\documentclass{article}

\usepackage{arxiv}
\usepackage[utf8]{inputenc} 
\usepackage[T1]{fontenc}    
\usepackage{hyperref}       
\usepackage{url}            
\usepackage{booktabs}       
\usepackage{amsfonts}       
\usepackage[numbers]{natbib}
\usepackage{amsmath, enumitem}
\usepackage{amssymb}
\usepackage{upgreek}
\usepackage{lineno}
\usepackage{soul}
\usepackage{fixltx2e}

\usepackage{nccmath}
\usepackage{ifthen}
\usepackage{color}
\usepackage{graphicx}
\usepackage{algorithm}
\usepackage{algorithmic}
\usepackage{setspace}
\usepackage{multirow}
\usepackage[T1]{fontenc}

\graphicspath{{Fig/}}

\title{Direct-3D Variational Bayesian Surface Wave Inversion and Its Application to Ambient Noise Tomography beneath Great Britain}
\date{} 					

\author{
	Xuebin Zhao \\
	School of Geosciences \\
	University of Edinburgh\\
	Edinburgh, Unite Kingdom \\
	\And
	Lily Irvin \\
	School of Geosciences \\
	University of Edinburgh\\
	Edinburgh, Unite Kingdom \\
	\And
	Erica Galetti \\
	School of Geosciences \\
	University of Edinburgh\\
	Edinburgh, Unite Kingdom \\
	\And
	Andrew Curtis \\
	School of Geosciences \\
	University of Edinburgh\\
	Edinburgh, Unite Kingdom \\
}

\begin{document}
	\maketitle

\begin{abstract}
We present a new, variational, fully nonlinear, probabilistic ambient noise tomography method, which estimates subsurface structure and quantifies the corresponding uncertainties directly in three dimensions (3D) from inter-receiver seismic surface wave dispersion data. We use the method to invert for high resolution 3D seismic velocity models of the upper crust beneath Great Britain using seismic ambient noise data recorded around the region -- a task that proved too high-dimensional and hence computationally demanding for Monte Carlo sampling to converge to a stable solution. We compare the inversion results from the new method to those obtained from two standard, indirect inversion methods, in which 2D (geographical) surface wave velocity maps and 1D (depth) shear velocity profiles are estimated in two separate, consecutive steps. The results show that the direct-3D scheme preserves better lateral continuity and produces better data fit than the two-step methods, and provides information about lateral correlations that is absent from the two-step solutions. The inversion results are consistent with large-scale geology of Great Britain, and for the first time provide seismologically-imaged evidence of the Great Glen Fault and other major tectonic faults. We therefore propose that direct-3D inversion schemes should be used where possible for surface wave inversion as they provide improved results at little additional computational cost.
\end{abstract}

\section{Introduction}
Seismic ambient noise tomography has emerged as a powerful tool for imaging the Earth's interior, particularly in seismically quiescent regions. Central to this approach is seismic interferometry, in which cross-correlation of long-duration, low-amplitude, ambient noise recordings (e.g., seismic signals generated by oceanic, atmospheric and human activities) between pairs of receivers allows the estimation of inter-receiver seismic wave seismograms. This process effectively converts receivers into virtual sources, generating an approximation to seismic data that would have been recorded at one of the two receivers if the other acted as an impulsive seismic source \cite{campillo2003long, snieder2004extracting, wapenaar2006green, curtis2006seismic, wapenaar2010tutorial1, wapenaar2010tutorial2, galetti2012generalised}. By analysing the retrieved seismograms, seismologists use ambient noise tomography to study the Earth's interior structure, even in regions with low levels of earthquake activity which would therefore otherwise be difficult using traditional earthquake tomography methods \cite{shapiro2005high, sabra2005surface, lin2007ambient, lin2008surface, bensen2008broadband, li2009ambient, de2011ambient, mordret2013near, allmark2018seismic, zhang20201, fone2024ambient}.

Great Britain is an island in the North Atlantic off the northwestern coast of continental Europe and is a typical region with low earthquake activity. A number of seismological studies have imaged the subsurface structure of the region using waves from earthquakes. For example, \citet{arrowsmith2005seismic} performed teleseismic body wave tomography to build a 3D P wave velocity model of the upper mantle beneath the British Isles; \citet{luckett2015local} created a seismic velocity model beneath Scotland using local earthquake data recorded over four decades; \citet{hardwick2009new} produced 3D seismic images of the crust beneath England, Wales and the Irish Sea using local earthquake tomography. However, earthquakes in and around great Britain are generally infrequent and of small magnitude \cite{baptie2010seismogenesis}; they provide limited data to image the subsurface. As an alternative, some studies imaged either deeper structures or sub-regions of the Earth beneath Ireland and Britain using data from active source-seismic refraction and reflection experiments \cite{klemperer1991deep, o1996lithosphere, o2010fine, o2012fine, chadwick1998seismic, masson1998wide, asencio2003mapping, landes2005review, kelly2007crustal, bastow2007spatial, maguire2011lispb, davis2012crustal, licciardi2020moho}. While these studies generally provide high resolution of crustal structures, they are expensive to conduct and so only span lines, or small patches of the geographical area of Great Britain. There have also been studies that attempt to image the UK using receiver functions \cite{champion2006crustal, tomlinson2006analysis, di2009nature, licciardi2014moho}, but this method only images the subsurface beneath the near-vicinity of receivers; unless receivers are placed densely in patches or along lines, this method therefore provides estimates of depth-profiles of significant seismic reflectors beneath isolated geographical points. 

Since Great Britain is surrounded by numerous seismic ambient noise sources from the Atlantic Ocean, the North Sea and the Norwegian Sea, it is reasonable to study this region using ambient noise tomography. \citet{nicolson2012seismic} presented the first Rayleigh wave group velocity map of the Scottish Highlands using ambient noise tomography, and \citet{nicolson2014rayleigh} constructed Rayleigh wave velocity maps across Great Britain and part of the island of Ireland. \citet{galetti2017transdimensional} cross-correlated the transverse component of the noise data to create Love wave group velocity maps across a similar region, and a 3D shear wave velocity model beneath the East Irish Sea by inverting the surface wave dispersion. \citet{zhao2022interrogating} conducted a similar inversion to estimate the 3D volume of the East Irish Sea basins. Deeper structures of the crust and upper mantle beneath Ireland and Britain were estimated from Rayleigh wave \cite{bonadio2021optimal} and Love wave \cite{chambers2023determining} phase velocity measurements, respectively. \citet{polat2012anisotropic} inverted for phase velocity maps and azimuthal anisotropy of Ireland's crust using Rayleigh wave information. 
Despite all of these studies, however, no current results provide high resolution 3D images of the upper crust beneath Great Britain that align well with known large-scale geology of the region. To fill this gap, in this study we perform surface wave inversion across the region using seismic ambient noise data.

In global seismology, the 3D Earth model is often parametrised by spherical harmonic basis functions (non-local representation), and their coefficients are estimated directly from surface wave dispersion data \cite[e.g.,][]{boschi2002new, kustowski2008anisotropic, ferreira2010robustness, chang2015joint}. However, particularly in regional or local ambient noise tomographic problems, 3D surface wave inversion is often solved via a two-step scheme to simplify the inversion. In this scheme, two-dimensional (2D) geographical maps of phase or group velocity surface wave speed at various frequencies (periods) are imaged first across the area of interest. These surface wave velocity maps are used to build dispersion curves at a grid of different locations, each of which is then inverted independently to estimate a shear wave velocity profile with depth beneath the corresponding spatial location \cite{trampert1995global, ritsema2020heterogeneity, snoke2002constraints, galetti2017transdimensional, bodin2009seismic, bodin2012transdimensionaltomo, mordret2013near, allmark2018seismic, crowder2021new, zhang20201, fone2024ambient}. 

A disadvantage of this `2D+1D' scheme is that results preserve only the limited information about lateral statistical correlations in the dispersion maps, because each 1D depth inversion is conducted independently of the others. In addition, while regularisation of 3D models can in principle be defined so as to encourage realistic 3D geological structures, this is not possible in the 2D dispersion inversion since we do not know, \textit{a priori}, what realistic group or phase velocity maps should look like. This leads to arbitrary regularisation in the two-step inversion schema \cite{nicolson2014rayleigh}. We note that results from previous two-step inversion studies (e.g., those cited above) sometimes show reasonable lateral continuity information. This is mainly because observed dispersion data themselves contain correct correlation information about subsurface structures (at least for nearby geographical locations), even if only part of that information is conveyed to the two-step solutions.

\citet{zhang20183} analysed the above issues and showed that due to the independence of many 1D inversions in the second step, and to possible errors introduced in the first step, the inversion results from the two-step method can be rather inaccurate. They therefore proposed a fully 3D inversion method that estimates the spatially 3D structure directly from frequency-dependent, inter-receiver dispersion measurements. The results showed that the new inversion method improves the accuracy of the velocity model estimation and produces more strongly laterally correlated, more intuitively reasonable uncertainties than the traditional two-step method \cite{zhang20201}. The method has since been tested and applied in a number of studies \cite[e.g.,]{roy2021accounting, kastle2025alpine}.

Since ambient noise surface wave inversion is nonlinear and solutions are nonunique, fully nonlinear Bayesian inversion is often applied to solve the problem probabilistically. This approach involves estimating the so-called \textit{posterior} probability density function (pdf) over model parameters. This describes the residual uncertainty in the model parameters after considering the observed data, and is considered to describe the complete solution to a Bayesian inverse problem. \cite{tarantola2005inverse}.

Markov chain Monte Carlo (McMC) methods are used widely to explore these uncertainties in surface wave tomography, by drawing random samples from the posterior pdf \cite{mosegaard1995monte, mosegaard2002monte, sambridge2002monte, sambridge2006trans}. However, these methods generally become computationally demanding for large scale inverse problems due to exponential growth in computation required to explore spaces of values of increasing numbers of parameters -- referred to as the curse of dimensionality \cite{curtis2001prior}. As an alternative, variational inference solves Bayesian inverse problems using numerical optimisation instead of random sampling. This offers greater efficiency and scalability in some high dimensional problems \cite{bishop2006pattern, blei2017variational}, and in previous work the method has been applied to increase the dimensionality of Bayesian seismic tomographic solutions \cite{zhang2019seismic, zhao2021bayesian, levy2022variational}.

In this study, we introduce a new, direct-3D method to perform Bayesian surface wave inversion, and apply it to estimate 3D seismic velocity structure beneath Great Britain. The dimensionality of the direct-3D inversion makes it almost impossibly expensive to solve using standard McMC methods. We therefore employ a recently introduced method called physically structured variational inference \cite[PSVI -][]{zhao2024physically} to reduce computational cost. We compare our inversion results to those from standard two-step nonlinear inversion schemes. We consider two commonly used parametrisations: regularly gridded models, and models tiled by Voronoi cells. We apply Metropolis-Hastings McMC (MH-McMC) and reversible jump McMC (rj-McMC) algorithms to these parametrisations, respectively, to perform the inversions. 

The main purpose of this work is to test and analyse advantages or disadvantages of the direct-3D inversion scheme compared to other related methods. We therefore perform tomography using an identical set of Love wave dispersion data to the previous study by \cite{galetti2017transdimensional}, who solved the problem using a Voronoi-cell based, two-step method. This allows direct comparison of results. In particularly, we show that the new method preserves stronger lateral correlations in the inversion results, thus providing higher inversion accuracy compared to two two-step schemes.



We first introduce Bayesian surface wave inversion using McMC and variational methods, followed by an overview of the standard two-step method, and then details of our direct-3D variational inversion scheme. We perform surface wave inversion across Great Britain using these methods, and compare results. Finally, we discuss our findings and draw conclusions.

\section{Methodology}
\subsection{Bayesian inversion}
Bayesian inference solves inverse problems probabilistically by evaluating the so-called \textit{posterior} probability density function (pdf) of model parameters $\mathbf{m}$ given observed data $\mathbf{d}_{obs}$ using Bayes' rule:
\begin{equation}
	p(\mathbf{m}|\mathbf{d}_{obs}) = \dfrac{p(\mathbf{d}_{obs}|\mathbf{m})p(\mathbf{m})}{p(\mathbf{d}_{obs})}
	\label{eq:bayes}
\end{equation}
Term $p(\mathbf{m})$ defines the \textit{prior} information about $\mathbf{m}$ that was available before inversion, and $p(\mathbf{d}_{obs}|\mathbf{m})$ is called the \textit{likelihood}, defined to be the probability of observing $\mathbf{d}_{obs}$ given any particular value of $\mathbf{m}$. Term $p(\mathbf{d}_{obs})$ is a normalisation constant called the \textit{evidence}.

Markov chain Monte Carlo (McMC) is commonly used to solve Bayesian inverse problems. In McMC, an ensemble of model realisations is sampled randomly from the posterior pdf in equation \ref{eq:bayes} \cite{tarantola2005inverse}. These model samples are used to approximate statistical properties of the posterior distribution, that characterise the set of possible solutions (the uncertainty). Theoretically, the density of McMC samples provides the \textit{correct} Bayesian posterior distribution as the number of samples tends to infinity \cite{brooks2011handbook}. Recently, a variety of Monte Carlo sampling algorithms have been developed to improve the sampling efficiency, for example by using gradient information to guide the sampling process \cite{fichtner2019hamiltonian, biswas2022transdimensional, zunino2023hmclab} or by incorporating advanced prior information into the inversion \cite{khoshkholgh2021informed, tsai2023towards, marignier2023posterior}. However, these Monte Carlo based methods typically require millions or more of forward simulations to reach reasonable convergence, making them highly computationally expensive. 

Variational inference is an alternative to Monte Carlo random sampling methods. This class of methods finds a probability distribution that best approximates the posterior pdf, out of a predefined family of probability distributions. The optimal pdf is determined by minimising the discrepancy between the variational approximation and the posterior distribution, typically measured by the Kullback-Leibler (KL) divergence \cite{kullback1951information}. As variational inference involves numerical optimisation rather than random sampling, it can be more efficient than McMC \cite{blei2017variational}.

Physically structured variational inference \cite[PSVI -][]{zhao2024physically} is an efficient variational method that employs a log-transformed Gaussian distribution to approximate the full, Bayesian posterior pdf. In PSVI, the Gaussian covariance matrix is structured specifically to capture only significant correlations in the posterior estimates of model parameter uncertainties. For example, in full waveform inversion (FWI) these are between pairs of parameters that are spatially proximal within a wavelength or so of each other, to account for the physical wavelength-averaging of the sensitivity of propagating waves to medium properties. This selective correlation pattern, ignoring all other parameter correlations, significantly reduces both the memory requirement and computational cost, while preserving the correlations that affect the solution most. The method has been applied to 2D FWI \cite{zhao2024variational}, 3D FWI \cite{zhao2025efficient} and time-lapse FWI \cite{zhao2025uncertainty} using synthetic data. These tests demonstrated that PSVI can deliver reasonable estimates of posterior uncertainties efficiently. In this study, for the first time, we apply PSVI to 3D surface wave inversion, and to real data.

\subsection{Two-step inversion scheme}
In the conventional two-step, 2D+1D scheme, spatially 2D surface wave velocity maps at various periods are first estimated from phase or group delays of surface waves travelling between source-receiver or receiver-receiver pairs. This is usually achieved by solving 2D travel time tomography problems at each period independently. At each grid node, a dispersion curve -- i.e., a profile of surface wave velocities that vary with period -- is constructed from the inversion results across various periods. In the second step, each local dispersion curve is inverted to estimate 1D shear wave velocity variations with depth beneath that location. These 1D depth inversions are usually run independently to allow perfect parallelisation. A 3D velocity model is constructed by concatenating the 1D inversion results.

Several issues arise with the two-step method. First, in the first step, additional prior information on surface wave velocities must be imposed to obtain a solution, which can bias the results of the shear wave velocity inversions \cite{nicolson2014rayleigh, zhang20201, bonadio2021optimal}. Second, since 1D depth inversions are usually conducted independently to enable parallelisation for computational efficiency, the final 3D results typically preserve little lateral correlation information. It is true that two-step inversion results can provide reasonable lateral continuity \cite[e.g.,][]{galetti2017transdimensional, fone2024ambient}, because dispersion data estimated from the first step are spatially correlated (at least for nearby geographical locations). However, this correlation is mainly due to the lack of resolution of the dispersed velocities at each location independently of others, and hence is again due to the regularisation required to stabilise solutions in the first step. Third, the two-step scheme yields 3D velocity models that fit the dispersion curves derived from the first step, rather than models that fit the original inter-receiver surface wave dispersion observations. This means that the inversion results may provide relatively poor data fits, as demonstrated in examples below. 

\subsection{Direct-3D variational inversion}
\citet{zhang20183} introduced a fully-3D Monte Carlo method that inverts for subsurface structures directly from inter-receiver surface wave dispersion data. This approach not only provides more accurate posterior model statistics, but also enhances lateral continuity compared to the two-step method. Since each velocity model is discretised in 3D, the dimensionality of the inverse problem is typically high, making it computationally expensive to solve using traditional McMC methods due to the curse of dimensionality. Trans-dimensional Bayesian inversion using rj-McMC was therefore used to reduce the number of unknown parameters to only those that are justifiably necessary to explain the data, improving sampling efficiency \cite{bodin2009seismic}.

In rj-McMC, subsurface velocity models are often parametrised parsimoniously, for example using Voronoi cells \cite{bodin2009seismic} or based on geological prior information \cite{guo2020bayesian}. Since a relatively low number of cells (hence unknown velocity parameters) is then used to parametrise a 3D velocity model, each cell generally represents a large region of the subsurface velocity model, within which a constant or smoothly varying velocity value is usually assigned \cite[e.g., Figure 8 in][]{bodin2009seismic}. Therefore, strongly laterally smoothed velocity models were obtained from previous studies \cite{galetti2017transdimensional, fone2024ambient}. In addition, the use of large cells can in itself lead to low posterior uncertainties in the inversion results, which may not reflect intra-cell velocity variations that are simply not well resolved by the data \cite{backus1968resolving}. A hierarchical Bayesian approach, often applied in conjunction with rj-McMC, treats data uncertainties as unknowns and adjusts them during Bayesian inversion \cite{bodin2012transdimensionaltomo}. However, this can sometimes affect the inversion accuracy. For example, \citet{galetti2017transdimensional} found that observational uncertainties tend to be upscaled (in that case by a factor of 2) to ensure that the posterior solution fit the data while remaining parsimonious (in other words, doubling the uncertainty was necessary in order to make the sparsely parametrised models consistent with the data in a Bayesian sense). Doubling the uncertainty means that some proposed samples whose synthetic data do not fit observed data to within their original, observationally estimated data uncertainties, might be accepted to within the inflated uncertainties. If data uncertainties were not adjusted during the inversion, many posterior samples would therefore have been rejected due to their low likelihood values. It is therefore expected that upscaled, artificially inflated data uncertainties can lead to posterior solutions of lower inversion accuracy, since data constraints (information) are thereby diminished. Finally, a recent study shows that both the trans-dimensional and hierarchical approaches introduce physical inconsistency into Bayesian inversion results: changes in model parametrisation lead to non-physical changes in the solution \cite{mosegaard2024inconsistency}.

To avoid these various issues, we perform direct-3D Bayesian surface wave inversion using fixed-dimensional regular grids of cells. Typically in rj-McMC, the shape of the Voronoi cells is updated to adapt to the shape of well resolved seismic velocity heterogeneities. These may or may not correspond to true geological boundaries depending on the lateral variation in resolution. To ensure that our regular gridded parametrisation can represent the true Earth model reasonably accurately, we use a small cell size, thus a large number of grid cells, to discretise a 3D velocity model. To manage the computational demands of solving the resulting high-dimensional inverse problem, we apply variational inference, specifically PSVI, instead of Monte Carlo. 

For forward simulation we use the two-step method of \citet{zhang20183}. The first step involves calculating surface wave velocities at periods of interest for any given 3D shear wave velocity model. This is achieved by extracting the 1D shear velocity profile beneath a dense grid of geographical locations, and calculating the dispersion curve corresponding to each of those 1D structures \cite{herrmann2013computer}. Subsequently, inter-receiver phase or group delay times at different periods are computed by solving the Eikonal equation using the fast marching method \cite{rawlinson2005fast}. Data-model gradients (used in PSVI) can be obtained by applying the chain rule to the gradients computed from these two separate steps.

\section{Setup of Tomography Problems}
\subsection{Data processing}
\begin{figure*}
	\centering\includegraphics[width=\textwidth]{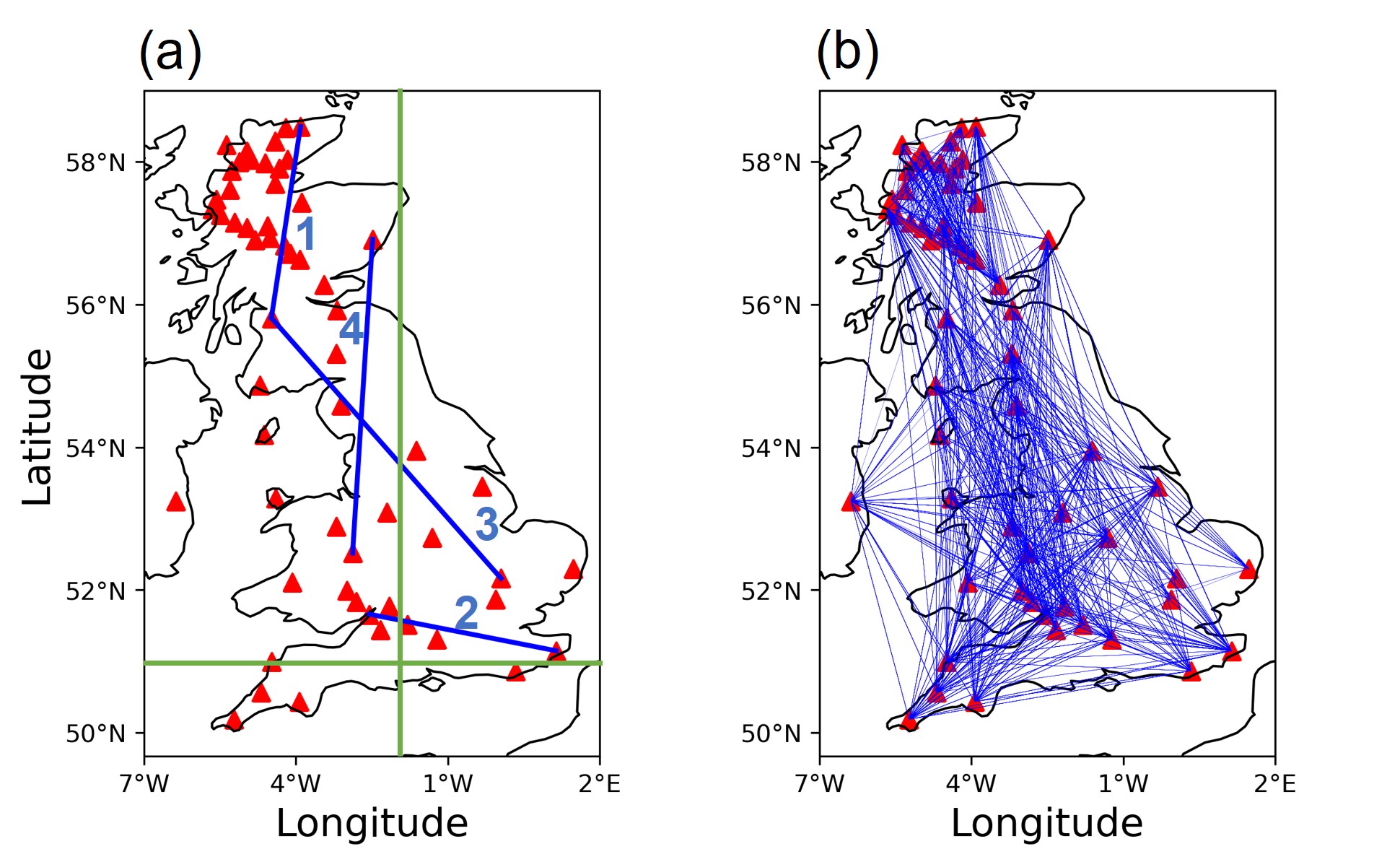}
	\caption{Locations of the seismometers (red triangles) used in this study. Thick blue lines in (a) represent 4 inter-receiver paths considered in Figures \ref{fig:uk_data_fit}. Green lines in (a) show locations of two vertical slices on which the inversion results are compared in Figure \ref{fig:uk_vertical}. Blue lines in (b) denote ray paths used to estimate Love wave dispersion data.}
	\label{fig:uk_stations_rays}
\end{figure*}

Figure \ref{fig:uk_stations_rays}a shows seismological stations (red triangles) installed around the Great Britain (one on Ireland) used in this study. The stations recorded a vertical (Z) component and two horizontal (north N, and east E) components of seismic ambient noise data in 2001-2003, 2006-2007 and in 2010. These data contain strong oceanic microseismic noise from the surrounding ocean and seas, and can be used to reconstruct the inter-receiver Green's functions. The vertical component was previously cross-correlated to construct inter-receiver Rayleigh wave dispersion data to perform Rayleigh wave tomography \cite{nicolson2014rayleigh}. While Rayleigh-wave tomography can show geological structures down to the lower crust and upper mantle, Love wave group velocity maps are expected to be more representative of shallow sedimentary and superficial layers. In this work we perform Love wave inversion using noise cross-correlations constructed from the two horizontal components by \citet{galetti2017transdimensional}.

During data processing, the noise recordings were first divided into 24-hour-long files, and then rotated into the transverse and radial directions. Cross-correlations of transverse day-files were then computed for considered station pairs (blue lines in Figure \ref{fig:uk_stations_rays}b) and stacked linearly over the total recording period to increase the signal-to-noise ratio. The positive time (causal) part and time-reversed negative time (acausal) part of the cross-correlation signals were stacked to obtain one-sided Green's function estimates for all considered inter-station paths. The multiple-filter analysis method \cite{herrmann2013computer} was used to estimate the fundamental mode Love wave group delays and hence path-averaged group velocities at periods of 4, 6, 8, 9, 10, 11, 12 and 15 s for the inter-station paths. In addition, daily cross-correlations were divided into several subsets, and the standard deviation values of the estimated period-dependent Love wave dispersion data across different subsets were used as the uncertainties of the dispersion data. Detailed station network information and a description of the data processing procedures can be found in \citet{galetti2017transdimensional}. In this test, we perform 3D surface wave inversion using exactly the same Love wave dispersion dataset from \citet{galetti2017transdimensional}, such that the inversion results obtained from the proposed direct-3D method can be compared fairly to those from \citet{galetti2017transdimensional}. 

\subsection{Inversion setup}
\label{sec:inversion_setup}

\begin{figure}
	\centering\includegraphics[width=\textwidth]{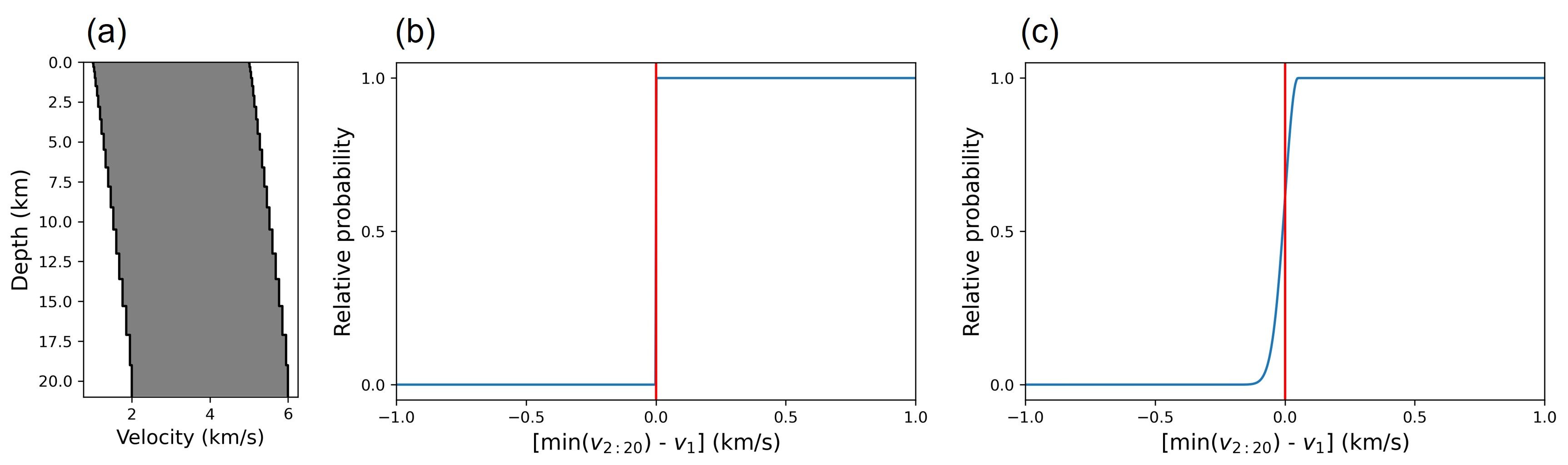}
	\caption{(a) Upper and lower bounds of the uniform prior distribution in different layers. (b) Prior pdf incorporating additional prior information in which shear wave velocity in the first layer is the lowest. This is used in the 1D Monte Carlo inversion. (c) Modified prior pdf used in the direct-3D variational inversion.}
	\label{fig:uk_prior_for_inversion}
\end{figure}

For both the direct-3D and two-step MH-McMC inversions, we discretise the imaged region, defined between longitudes 7$^\circ$W and 2$^\circ$E and latitudes 49.67$^\circ$N and 59$^\circ$N, into a regular grid of 28$\times$29 cells with a spacing of 0.33$^\circ$ in both the latitude and longitude directions. As shown in Figure \ref{fig:uk_prior_for_inversion}a, shear wave velocity structure is parametrised from the surface down to a depth of 20 km across 20 layers. This depth range was chosen based on a sensitivity analysis of Love wave group velocities for periods ranging from 4 to 15 s. Layer thicknesses increase incrementally from 0.1 km in the first layer to 2 km in the last layer, with an increase of 0.1 km in each subsequent layer. This is designed to reflect the decreasing sensitivity of surface waves at greater depths. This stack of layers is underlain by an elastic half-space.

We define a uniform prior distribution for shear wave velocities at different depths, setting broad interval bounds to ensure that the final inversion results are not biased by the prior limitations \cite{galetti2017transdimensional}. The lower and upper bounds of these intervals are displayed in Figure \ref{fig:uk_prior_for_inversion}a. We include additional prior information specifying that shear wave velocity in the first layer is the lowest, to ensure that modelled dispersion curves represent the fundamental mode \cite{herrmann2013computer}. This setup is illustrated in Figure \ref{fig:uk_prior_for_inversion}b, and can be implemented easily in the 1D Monte Carlo inversion (the two-step scheme), by rejecting samples in which velocity values in the first layer are not the lowest. However, incorporating this condition into the one-step variational inversion is less straightforward since the method requires the gradient of the prior probability density value to be evaluated, and the prior pdf in Figure \ref{fig:uk_prior_for_inversion}b is not differentiable at a zero value. To address this, we use a modified prior pdf displayed in Figure \ref{fig:uk_prior_for_inversion}c, in which a Gaussian averaging filter is introduced around zero to ensure that the gradient of the prior distribution can be calculated easily. We admit that for some cases the velocity in the top layer is not the lowest and the corresponding prior probability value is thus not strictly zero from Figure \ref{fig:uk_prior_for_inversion}c. Nevertheless, in our preliminary tests we performed 1D Monte Carlo inversions using each of the two prior distributions, in which a small standard deviation value is set for the Gaussian filter. We obtained almost identical posterior solutions, proving that the effect from the modified prior distribution is negligible.

We employ PSVI for the direct-3D inversion, and the likelihood function is defined to be an uncorrelated Gaussian distribution in data space, using the data standard deviations estimated as explained above. For the two-step MH-McMC inversion, we use PSVI in the first step (2D lateral surface wave tomography at each period) due to the dimensionality of the problem. Again, we define a non-informative uniform prior pdf bounded between 1-6 km/s for the Love wave group velocities at all considered periods, to ensure that this prior does not bias the final inversion results. The second step is solved using the MH-McMC algorithm. This is because MH-McMC should provide the correct Bayesian posterior solution to an inverse problem once the Markov chains have converged, and given the relatively low dimensionality (20 parameters) of each depth inversion, the computational cost of running the algorithm to the point of reasonable convergence is not unattainable. To remove any potential bias caused by using MH-McMC instead of PSVI in the second step inversion, we also apply PSVI for the depth inversion and provide the corresponding inversion results in Appendix \ref{ap:inversion_results}. This essentially provides results from a separate method, although one that clearly is not independent of the Monte Carlo two-step method since they both use the same dispersion data from the first step. In the rest of this paper we do not explicitly compare and analyse the results from the two-step PSVI inversion with those obtained from other methods, since the two-step PSVI results are used only to validate the two-step MH-McMC results, and the two sets of results are highly consistent.

To summarise, exactly the same hyper-parameters are defined to perform the direct-3D and the two-step MH-McMC inversions to ensure that differences between the two sets of inversion results are attributed to the methods themselves. In particular, the same regular gridded discretisation, two-step forward simulation method, Gaussian likelihood function, and uniform prior distribution are used (other than the small difference illustrated in Figures \ref{fig:uk_prior_for_inversion}b,c). 

The two-step rj-McMC inversion was reported in \citet{galetti2017transdimensional}. The inversion setup is almost the same as that for the two-step MH-McMC inversion, the main differences being the use of trans-dimensional and hierarchical inversion approaches \cite{bodin2012transdimensionaltomo}.

In the next section, we compare three sets of inversion results obtained from the direct-3D PSVI, two-step MH-McMC (which uses PSVI and MH-McMC in the two steps) and the two-step rj-McMC \cite[from][]{galetti2017transdimensional} methods.

\section{Results}
\subsection{Inversion results}

\begin{figure}
	\centering\includegraphics[width=\textwidth]{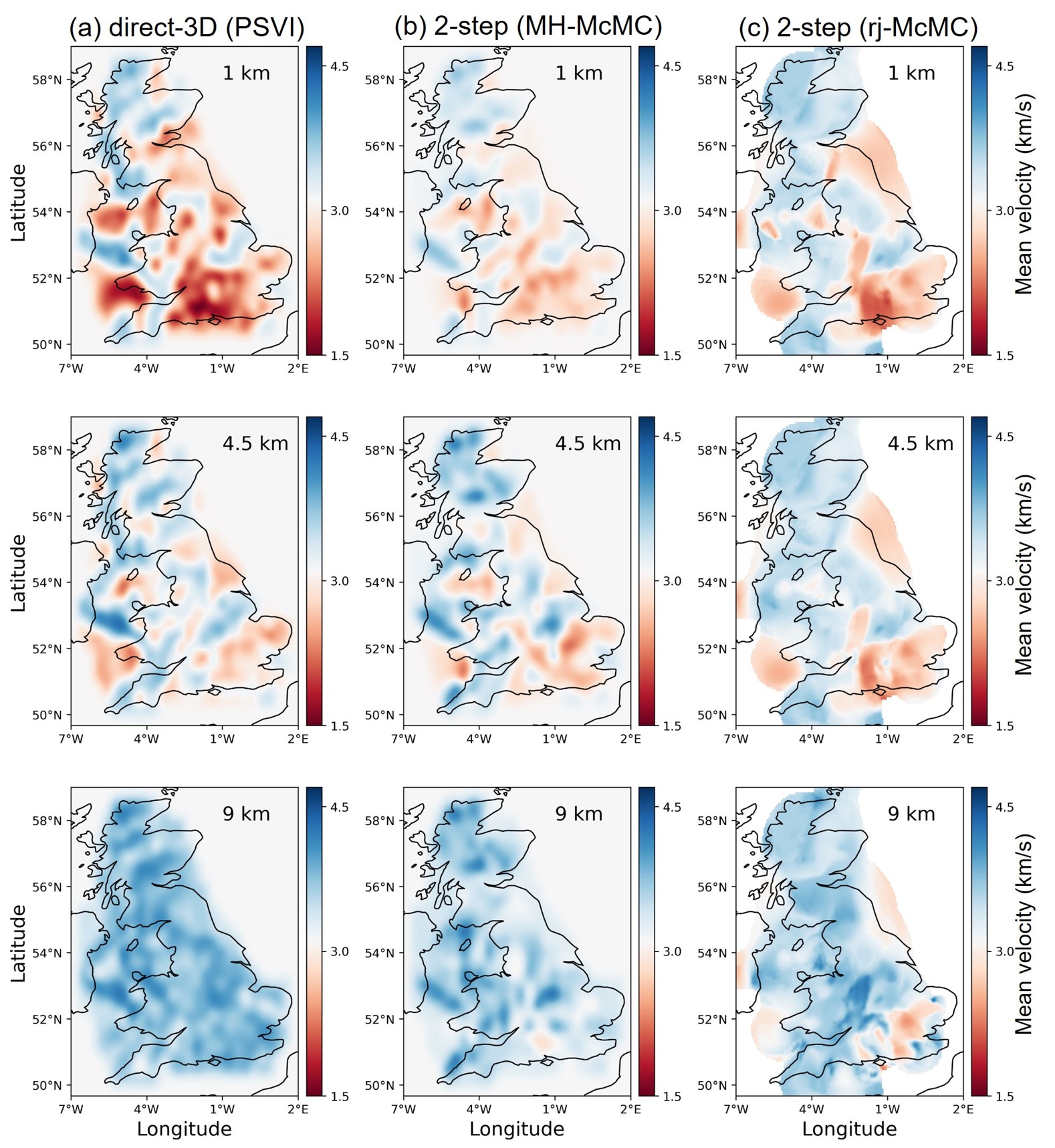}
	\caption{Horizontal slices of the inverted mean velocity maps using three inversion methods, at depths of 1 km, 4.5 km and 9 km.}
	\label{fig:uk_horizontal_mean_3methods}
\end{figure}

\begin{figure}
	\centering\includegraphics[width=\textwidth]{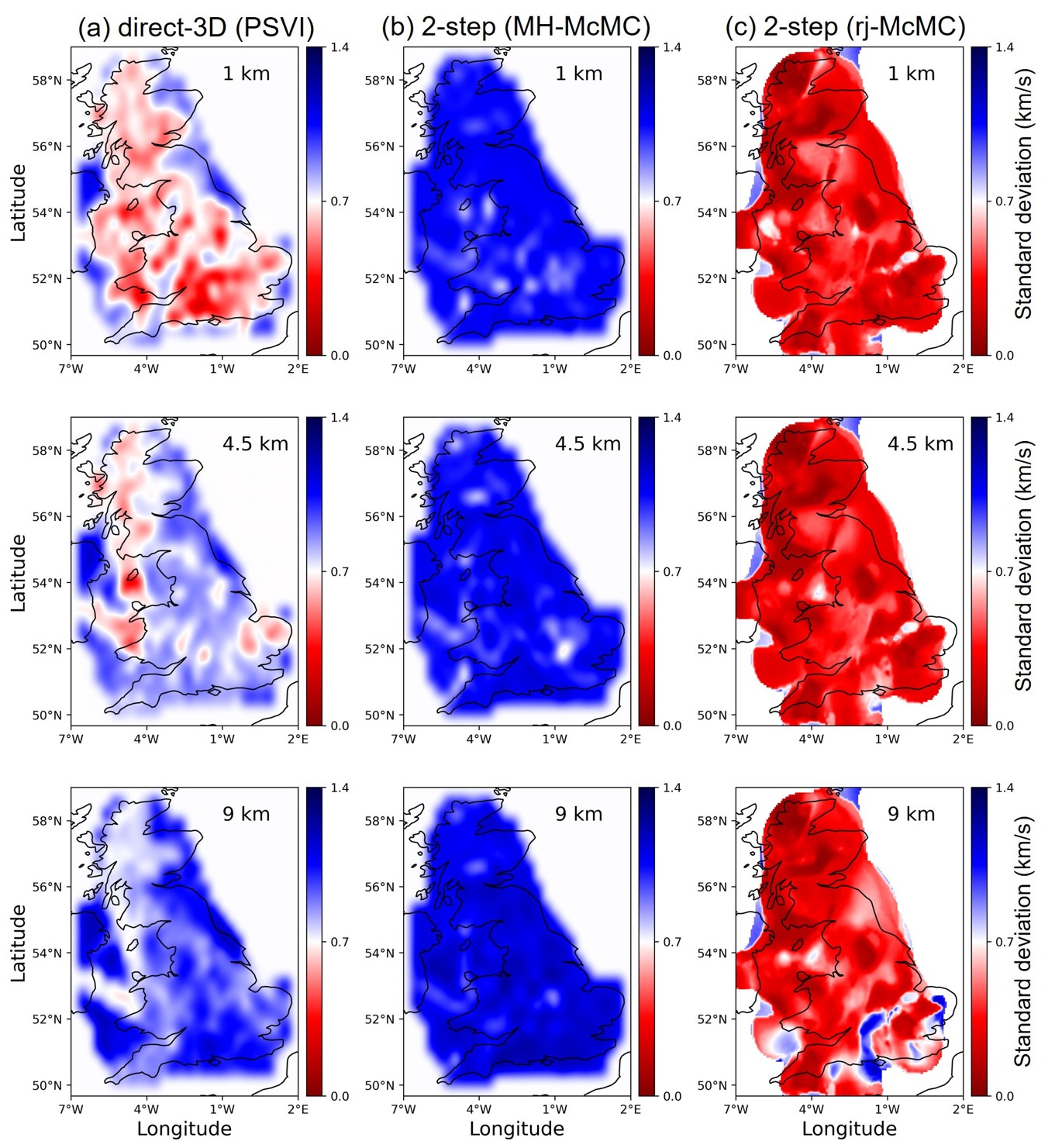}
	\caption{Standard deviation maps associated with the mean velocity maps in Figure \ref{fig:uk_horizontal_mean_3methods}. Note that the same color scale is applied to all panels to highlight the relative amplitude of standard deviation values from the 3 sets of results.}
	\label{fig:uk_horizontal_std_3methods_samescale}
\end{figure}

Figures \ref{fig:uk_horizontal_mean_3methods} and \ref{fig:uk_horizontal_std_3methods_samescale} display the posterior mean and standard deviation maps of 3 horizontal slices at depths of 1 km, 4.5 km and 9 km, respectively, obtained using the three inversion methods. For better comparison, we show only regions that are covered by dense ray paths (Figure \ref{fig:uk_stations_rays}b). The same color scale is applied to all panels in Figure \ref{fig:uk_horizontal_std_3methods_samescale} to highlight the relative amplitude of standard deviation values from the 3 sets of results. In Appendix \ref{ap:inversion_results} Figures \ref{fig:uk_horizontal_std_3methods} and \ref{fig:uk_vertical_diffscale}, we provide the corresponding maps using different color scales for each panel to highlight the different uncertainty structures. We also display several posterior samples from the direct-3D inversion in Appendix \ref{ap:inversion_results}, Figures \ref{fig:uk_horizontal_3d_bestfit_samples} and \ref{fig:uk_horizontal_3d_worstfit_samples}.

At the largest scale of low and high velocity regions, the three sets of mean velocity maps in Figure \ref{fig:uk_horizontal_mean_3methods} are roughly consistent. However, details in the left two columns differ from those in the right column, differences which may arise from the use of different parametrisations: regular grid cells on the left and centre versus Voronoi cells on the right, which implicitly impose different strengths of smoothing in the prior information \cite{bodin2009seismic}.

In Figure \ref{fig:uk_horizontal_std_3methods_samescale}, the overall standard deviation depicted from the direct-3D inversion is lower than that from the two-step MH-McMC inversion. This may occur because, first, we use PSVI for the fully 3D inversion and use MH-McMC for the 1D inversion. Previous studies indicated that PSVI tends to underestimate posterior uncertainties due to its Gaussian based assumption to the posterior pdf \cite{zhao2024physically} (this is verified by Figures \ref{fig:uk_horizontal_psvi_twostep} and \ref{fig:uk_vertical_psvi_twostep} in Appendix \ref{ap:inversion_results} in which the two-step PSVI inversion is performed). Second, the dimensionality of the 3D inversion is significantly higher than that of each 1D inversion (16,240 versus 20). The curse of dimensionality could lead to a phenomenon known as mode collapse, in which posterior uncertainties are strongly underestimated in high dimensional inverse problems. Third, as is standard practice in the two-step inversion, although we obtain full posterior pdfs of the Love wave group velocity models at the periods considered from the first step, only the resulting mean and standard deviation values are used in the second step. This results in the loss of other statistical information -- such as spatial correlations between group velocity values at different locations -- which is therefore not used to constrain the second step of the inversion. The absence of these correlations generally results in higher (overestimated) uncertainties. We also note that the posterior uncertainties from the two-step rj-McMC are lower than those from the other two methods which is therefore primarily due to use of the Voronoi cells. 

\begin{figure*}
	\centering\includegraphics[width=\textwidth]{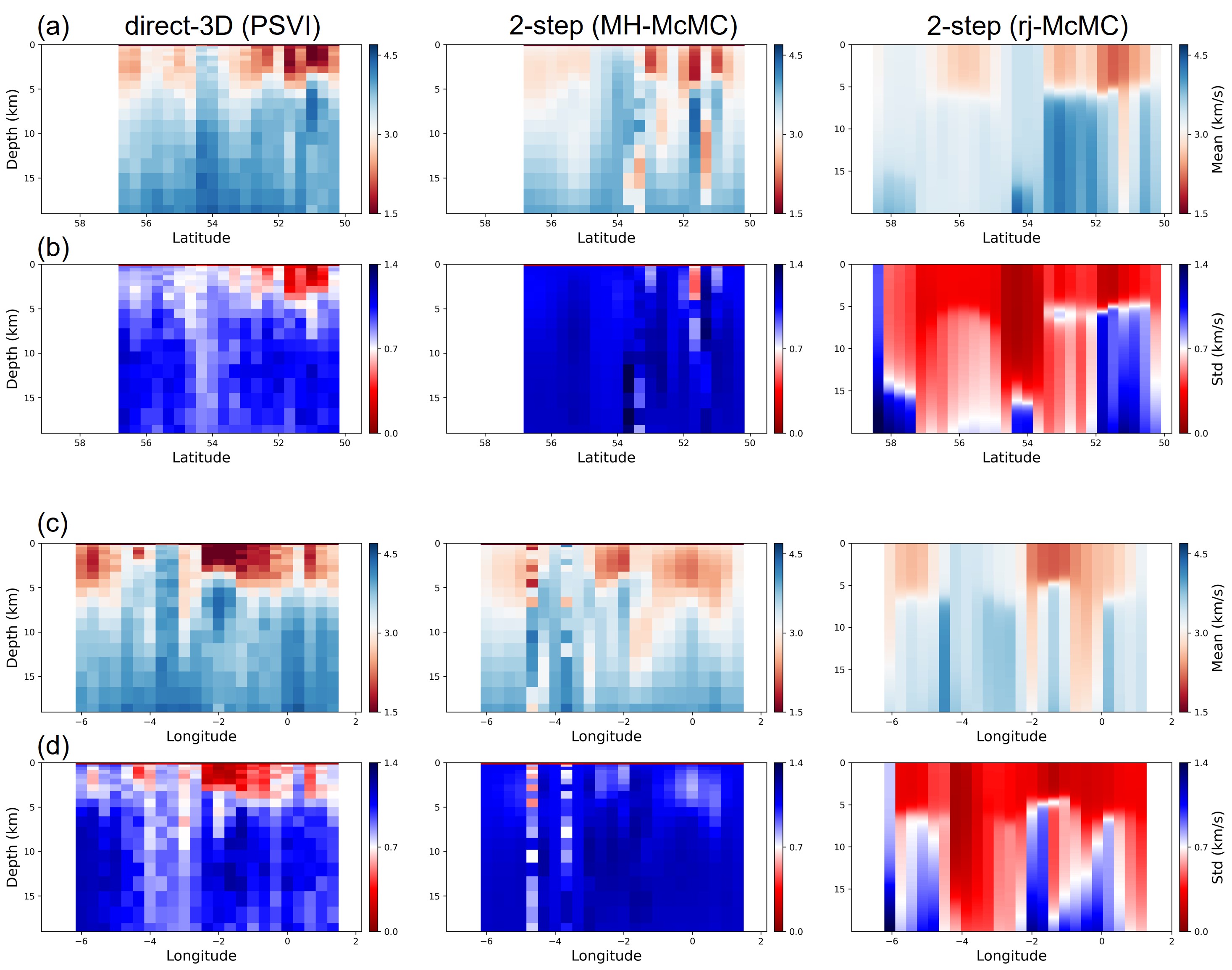}
	\caption{(a) and (c) Average velocity, and (b) and (d) standard deviation maps for two vertical slices through the inversion results at (a) and (b) 2$^\circ$W longitude, and (c) and (d) 51$^\circ$N latitude, respectively. The locations of these slices are marked by the green lines in Figure \ref{fig:uk_stations_rays}a.}
	\label{fig:uk_vertical}
\end{figure*}

In Figure \ref{fig:uk_vertical}, we compare two vertical slices extracted from the three sets of results; their locations -- at 51$^\circ$N latitude and 2$^\circ$W longitude -- are marked by two green lines in Figure \ref{fig:uk_stations_rays}a. Regions that are barely updated by the inversion, thus retaining posterior statistics that are nearly identical to the prior statistics, have been omitted.

The two-step inversion results exhibit limited lateral correlation (continuity). In Figure \ref{fig:uk_vertical}, both the mean and standard deviation maps from the 2-step MH-McMC and rj-McMC inversions display far more laterally-sharp discontinuities across the sections; the vertical slices from the direct-3D inversion show better lateral continuity. In Figure \ref{fig:uk_vertical} we project the original two-step rj-McMC results in \citet{galetti2017transdimensional} onto the same geographical grid as that used in the other two methods, so that the three results can be compared at the same spatial resolution. However, previous studies show that the two-step rj-McMC inversion results preserve little lateral correlation information, even when projected onto a much finer grid \cite{zhang20201}. Note that in both the direct-3D and two-step MH-McMC inversions, we define additional prior information in which the top layer (which has a thickness of 100 m from Figure \ref{fig:uk_prior_for_inversion}a) has the lowest velocity value. Therefore, in the left two panels in Figures \ref{fig:uk_vertical}a and \ref{fig:uk_vertical}c we observe a low velocity layer (dark red color) as a very thin line due to its thickness of 0.1 km.

We also display the two-step inversion results obtained using PSVI for both two steps in Appendix \ref{ap:inversion_results}, Figures \ref{fig:uk_horizontal_psvi_twostep} and \ref{fig:uk_vertical_psvi_twostep}. The results are similar to those from the two-step MH-McMC method, proving that the main differences in the inversion results from the direct-3D and two-step MH-McMC methods observed in Figures \ref{fig:uk_horizontal_mean_3methods} and \ref{fig:uk_vertical} are caused by splitting the inversion into two separate steps, and not from the difference between Monte Carlo and variational inference.

We compared the accuracy of inversion results obtained using the direct-3D and two-step MH-McMC methods by generating 1000 random model samples from the corresponding posterior pdfs, and calculating synthetic inter-receiver Love wave group delay times for each model. Note that the two-step MH-McMC inversion does not provide posterior samples of 3D shear velocity models explicitly, but rather independent random samples from each 1D profile. We therefore randomly select samples of each 1D velocity profiles from each geographical location and concatenate them to form composite 3D model samples. We are unfortunately unable to draw samples from the rj-McMC results from the information provided in \citet{galetti2017transdimensional}.

\begin{figure*}
	\centering\includegraphics[width=\textwidth]{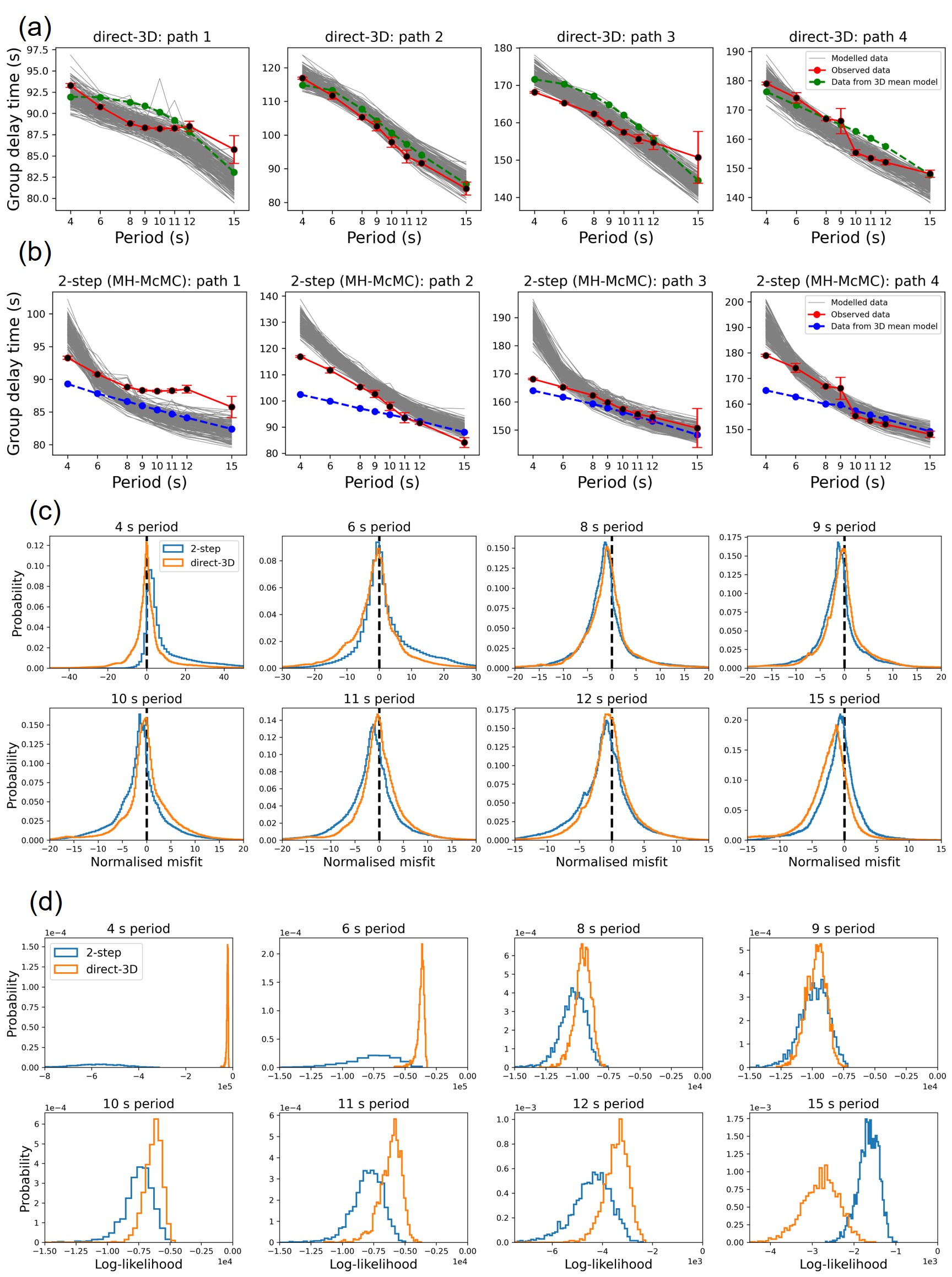}
	\caption{Observed (red lines) and synthetic (grey lines) Love wave group delay times using posterior samples from (a) the direct-3D inversion and (b) the two-step MH-McMC inversion, respectively, between 4 inter-receiver paths (columns) denoted by blue lines in Figure \ref{fig:uk_stations_rays}a. Dashed green and blue lines show data simulated from the posterior mean velocity models in each case. Histograms of (c) normalised data misfit values and (d) logarithmic likelihood values obtained from the two sets of inversion results. The normalised misfit value is the difference between the observed and synthetic data, divided by the data standard deviation at each data point, and the logarithmic likelihood value is obtained by summing the squared normalised misfit values across all data points and multiplying by -0.5 (according to the definition of a Gaussian likelihood function).}
	\label{fig:uk_data_fit}
\end{figure*}

Grey lines in Figures \ref{fig:uk_data_fit}a and \ref{fig:uk_data_fit}b show modelled Love wave dispersion data (in this case group delay times) using posterior samples from the direct-3D and two-step MH-McMC inversion methods, respectively, between 4 representative inter-receiver paths denoted by 4 blue lines in Figure \ref{fig:uk_stations_rays}a. Since our observed data are Love wave group delay times of waves propagating between pairs of stations, we observe that surface wave propagation time decreases (average inter-station velocity increases) with increasing period. Red lines and error bars represent the corresponding observed data and data uncertainties, respectively. The magnitude of the observed Love wave group delay time data is large, so the standard deviation error bars from Figure \ref{fig:uk_data_fit} appear to be very small -- on average, data uncertainties are around 1\% of the observed data. 

Generally, the modelled data from the direct-3D results in Figure \ref{fig:uk_data_fit}a fit the observed data better compared to those from the two-step MH-McMC method in Figure \ref{fig:uk_data_fit}b. This is primarily because the two-step inversion only fits dispersion curves derived from the first step, not the actual observed inter-receiver Love wave group delay times shown in the figure. Dashed green and blue lines in Figures \ref{fig:uk_data_fit}a and \ref{fig:uk_data_fit}b show data modelled from the posterior mean velocity model from each of the two methods, and again support the results above: the green lines from the direct-3D inversion in Figure \ref{fig:uk_data_fit}a align more closely with the red observed data. Further comparisons involving additional inter-receiver paths are presented in Appendix \ref{ap:inversion_results}, Figures \ref{fig:uk_16paths}, \ref{fig:uk_16paths_data_3d} and \ref{fig:uk_16paths_data_1d}. Another intriguing observation is that neither the dashed green nor blue lines align perfectly with the mean of the corresponding grey lines in each figure, with the deviation being more obvious for the dashed blue line. This discrepancy is caused by the nonlinearity of the forward problem since in a perfectly linear scenario these dashed lines would align precisely with the mean of the grey lines. 

In Figure \ref{fig:uk_data_fit}b, we observe that the synthetic data simulated from the posterior samples (grey lines) are larger than the observed data, whereas those from the posterior mean model (dashed blue lines) are smaller than the observed data, which seems slightly unusual. However, we need to take into account that in the two-step method, the posterior samples are obtained only by fitting the dispersion curves extracted from the first step (travel time tomography), rather than the original inter-receiver group delay times (the red lines). In other words, the red lines are not the \textit{actual data} used to generate posterior samples in the two-step inversion. Therefore, there is no reason that either the grey or dashed blue lines \textit{should} fit (at least closely) the red lines. This is a primary disadvantage of two-step methods, and highlights the necessity of using the direct-3D inversion method for 3D surface wave inversion. 



We also calculate normalised data misfit values for the two sets of posterior samples. The normalised data misfit is defined here as the difference between synthetic and observed data relative to data uncertainty (i.e., divided by standard deviation) at each data point. Figure \ref{fig:uk_data_fit}c displays histograms of the normalised data misfit values at 8 periods, with zero misfit indicated by a dashed black line in each figure. Overall, the orange histograms from the direct-3D method display a central concentration closer to the zero misfit value, except for results at 15 s period. Given our use of an uncorrelated Gaussian likelihood function for Bayesian inversion, we further calculate the logarithmic likelihood values for each posterior sample. Figure \ref{fig:uk_data_fit}d shows the corresponding histograms. From Figures \ref{fig:uk_data_fit}c and \ref{fig:uk_data_fit}d, we observe that the misfit values from the direct-3D inversion are consistently lower, and the likelihood values are higher, compared to those from the two-step MH-McMC inversion. Therefore, we conclude that samples from the direct-3D inversion method yield better fits to the observed data than those from the two-step MH-McMC method.

To confirm that our observations above are more broadly applicable to surface wave inversion problems, we compare the direct-3D and two-step MH-McMC inversion methods using an additional synthetic checkerboard example, which mirrors the setup of the real data example; in other words, everything in the tomographic problem setup is the same as that used in the real data example above, the only difference being between the \textit{true} velocity model in each case. More details about this example are provided in Appendix \ref{ap:synthetic}.

\begin{figure*}
	\centering\includegraphics[width=\textwidth]{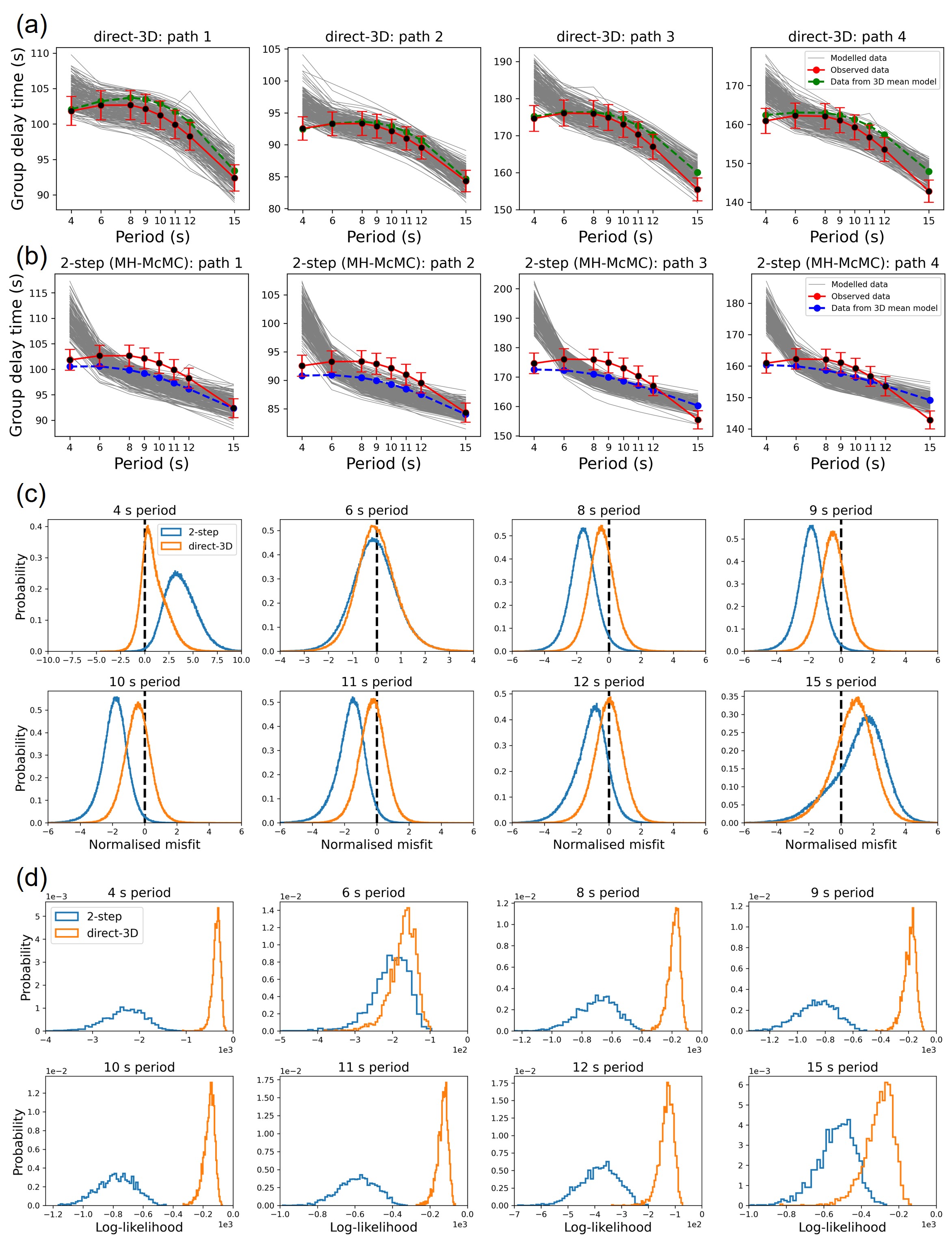}
	\caption{Comparison of data fitting for a synthetic checkerboard example. Key as in Figure \ref{fig:uk_data_fit}.}
	\label{fig:synthetic_data_fit}
\end{figure*}

Figure \ref{fig:synthetic_data_fit} compares the synthetic Love wave group delay times, normalised data misfit, and the logarithmic likelihood values obtained in this synthetic example using posterior samples from the two methods, similarly to those displayed in Figure \ref{fig:uk_data_fit}. This again demonstrates that the inversion results from the direct-3D method are more accurate than those from the two-step MH-McMC method, reaffirming that the conclusions drawn above may be more broadly applicable to other seismic surface wave inversion problems. 

These two examples suggest that the use of two-step methods for surface wave tomography introduces significant and avoidable artefacts. If seismologists instead use the direct-3D inversion methods herein, this would preserve lateral correlations discussed in previous studies \cite{zhang20183}, but more importantly, would provide significantly better data fits and more accurate inversion results.

\subsection{Interpretation}
\begin{figure*}
	\centering\includegraphics[width=\textwidth]{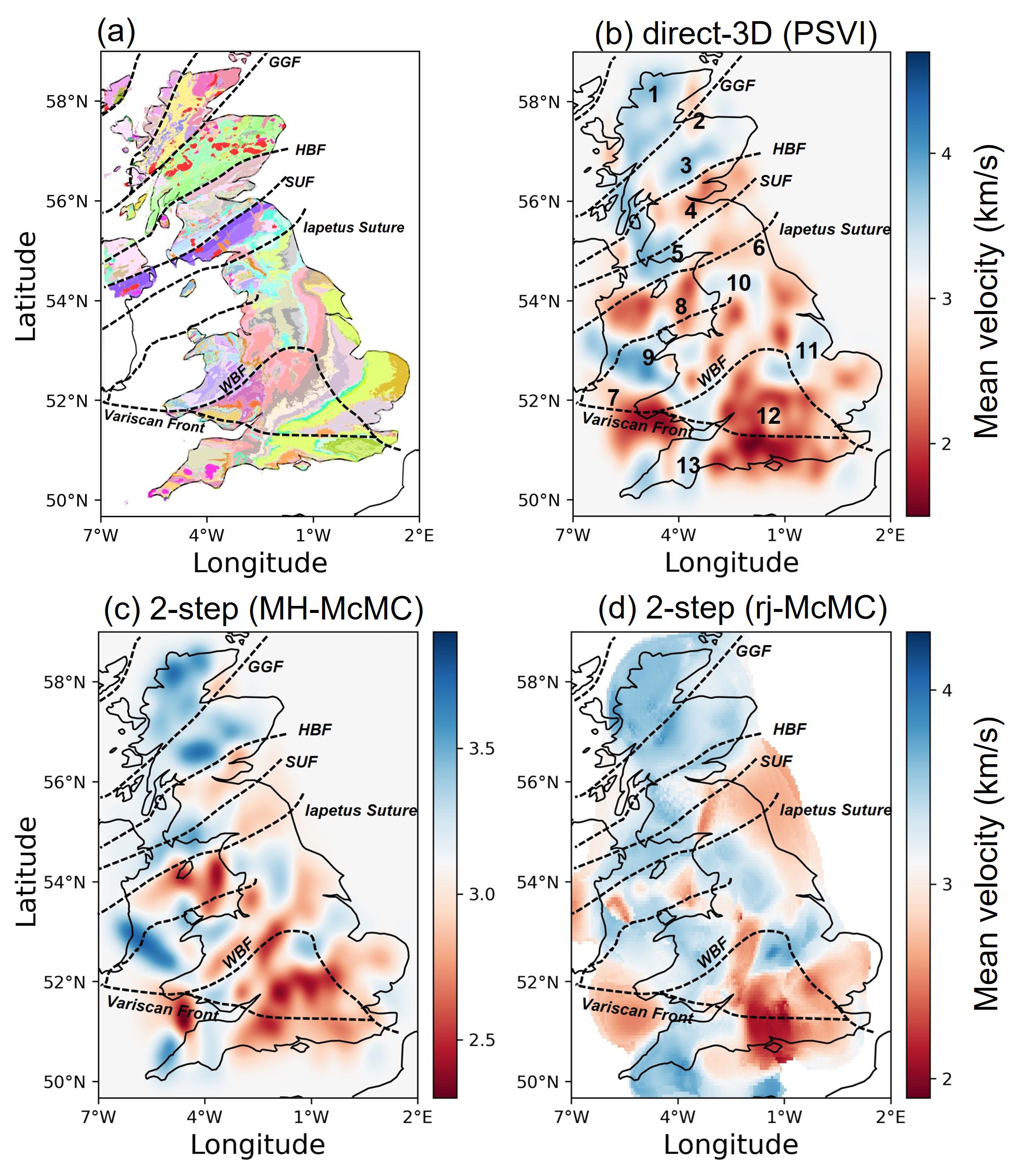}
	\caption{(a) Bedrock geology map of Great Britain with detailed legends reported in the British Geological Survey GeoIndex Onshore \cite{BGS2020Geo}. (b), (c) and (d) Mean velocity maps at 1 km depth obtained from the 3 inversion methods. Black numbers in (b) are used to mark main geological units with distinct terranes and to denote specific locations discussed in the main text. Major faults corresponding to terrane boundaries are abbreviated as follows: Great Glen Fault (GGF); Highland Boundary Fault (HBF); Southern Uplands Fault (SUF); Welsh Borderland Fault System (WBF). Different color scales are applied in panels (b), (c) and (d) for better comparison.}
	\label{fig:uk_interpretation}
\end{figure*}

We validate the inversion results by comparing them to the known geology. Figure \ref{fig:uk_interpretation}a shows the bedrock geological map of the studied area, obtained from the British Geological Survey GeoIndex Onshore \cite{BGS2020Geo}. Dashed black lines mark major terrane boundaries around the region. Figures \ref{fig:uk_interpretation}b, c and d display posterior mean velocity models at 1 km depth obtained using the three methods. Color scales in these panels are optimised individually to ensure that they can highlight the overall geological features of the studied region clearly. 

Generally in Figure \ref{fig:uk_interpretation}, the inverted mean maps from the three tested methods show reasonable spatial correlation with the major structures on the geological map (Figure \ref{fig:uk_interpretation}a) and previous tomographic studies of Great Britain \cite[e.g.,][]{nicolson2012seismic, nicolson2014rayleigh, galetti2017transdimensional, bonadio2021optimal}. They also provide some interesting insights into the 3D structure of the crust under Great Britain. High velocities are observed around the Scottish Highlands of Northern Britain due to its metamorphic and igneous origin \cite{trewin2002geology}. Within this region we observe some spatial variations of velocities between different groups. For example, fast velocity anomalies in the Northern Highlands (annotation 1 in Figure \ref{fig:uk_interpretation}b; hereafter we use the number of each annotation to denote its corresponding location for simplicity) are coincident with the old Lewisian rock of the Hebridean terrane and can also be associated with gravity and magnetic anomalies \cite{nicolson2012seismic}. The low velocities close to the north east coast (annotation 2 in Figure \ref{fig:uk_interpretation}b) can likely be attributed to the thick, sedimentary pull-apart basin in the Moray Firth and to the Devonian sediments situated along the north east coast \cite{trewin2002geology}. This is observed clearly in Figures \ref{fig:uk_interpretation}b and \ref{fig:uk_interpretation}c (although it is more evident in the former) and vaguely in Figure \ref{fig:uk_interpretation}d. 

In Figure \ref{fig:uk_interpretation}b, we observe a slightly low velocity region along the Great Glen Fault (GGF) between the following two high velocity regions: the Northern Highlands (1 in Figure \ref{fig:uk_interpretation}b) and the Central Highlands (3 in Figure \ref{fig:uk_interpretation}b), extending northwards into the Northern Highlands terrane and tracking along the Glenfinnan group, where most of the Lewisian inliers within the northern terrane are located. This feature is slightly less obvious from the two-step MH-McMC results (Figure \ref{fig:uk_interpretation}c), and is almost invisible from the two-step rj-McMC results (Figure \ref{fig:uk_interpretation}d), and in previous seismic imaging studies around the British Isles \cite{nicolson2014rayleigh, galetti2017transdimensional}. The Great Glen Fault is observable down to 9 km depth from the direct-3D inversion results in Figure \ref{fig:uk_horizontal_mean_3methods}a.

Around annotation 4 in Figure \ref{fig:uk_interpretation}b, there is a low velocity zone located south of the Highland Boundary Fault and north of the Southern Uplands Fault (SUF), corresponding to the Devonian and Carboniferous rocks that form the Midland Valley sedimentary basin. In Figure \ref{fig:uk_horizontal_mean_3methods}, the low velocity body can be observed down to about 4.5 km depth and is invisible by 9 km depth. This is consistent with previous studies \cite{dentith1989mavis, dentith1990mavis}, in which approximately 4-8 km of sediment were found from seismic refraction studies across the Midland Valley. This is not so clearly visible in the two-step rj-McMC inversion results. 

A low velocity anomaly is observed beneath the East Irish Sea (8 in Figure \ref{fig:uk_interpretation}b), corresponding to several sedimentary basins such as the Lagman Basin and the Eubonia Basin. These features are consistent with \citet{zhao2022interrogating}, in which 3D surface wave inversion beneath the East Irish Sea was performed using rj-McMC. They used the uncertainty results to estimate basin area in the near surface, which gave results close to the results estimated from surface geological surveys \cite{mellett2015geology}. Another two low velocities can also be observed, one of which is located at the offshore sedimentary basins along the east coastline of mainland Britain from 3$^\circ$W, 56$^\circ$N to 0$^\circ$E, 54$^\circ$N (6 in Figure \ref{fig:uk_interpretation}b) and the other from East Ireland to Southwest Wales (7 in Figure \ref{fig:uk_interpretation}b). 

The Midland Platform is a region of relatively undeformed Precambrian basement, corresponding to an area of low velocity (12 in Figure \ref{fig:uk_interpretation}b), which spans several sedimentary basins such as the Cheshire Basin (2.5$^\circ$W, 52.5$^\circ$N), the Anglian-London Basin (0$^\circ$W, 52$^\circ$N), the Weald Basin (0$^\circ$W, 51$^\circ$N) and the Wessex Basin (3$^\circ$W, 50.5$^\circ$N). Previous research found that the Midland Platform is a region of high crustal thickness and that average crustal velocities here are lower than in the surrounding region \cite{chadwick1998seismic, tomlinson2006analysis, hardwick2009new}. All three sets of results in Figures \ref{fig:uk_interpretation}b, c and d recovered this large low velocity anomaly clearly. In addition, gravity data in this region show a clear negative Bouger anomaly that could also be associated with thick crust \cite{nicolson2014rayleigh}.

Around the Southern Uplands, a SW-NE trending high velocity area is observed  (5 in Figure \ref{fig:uk_interpretation}b) and can be attributed to the siltstones, wackes and felsic plutons of the Southern Uplands accretionary complex. Several high velocity anomalies are observed surrounding the East Irish Sea basins including the Northwest Wales (9 in Figure \ref{fig:uk_interpretation}b) and the Lake District in the Northwest England (10 in Figure \ref{fig:uk_interpretation}b), which are also observed in previous studies \cite{nicolson2014rayleigh, galetti2017transdimensional, bonadio2021optimal}. The high velocity region to the Northeast of the Midland Platform (11 in Figure \ref{fig:uk_interpretation}b) appears to correlate with the northern limit of the Anglo-Brabant Massif \cite{pharaoh2018anglo} in the Eastern England, and is colocated with a series of positive magnetic anomalies and a high gravity structure \cite{nicolson2014rayleigh}. High velocities in the Cornwall area (13 in Figure \ref{fig:uk_interpretation}b) can be attributed to the intrusion of granite into the surrounding sedimentary rocks formed during the Variscan orogeny. 


\section{Discussion}
We first discuss the computational cost of the three inversion methods tested. For the direct-3D method, we perform variational Bayesian inversion with 15,000 iterations, and use 8 samples per iteration to estimate the data-model gradient. This results in a total of 120,000 forward simulations. For the two-step MH-McMC method, 2D variational dispersion tomography is performed at each period with 10,000 iterations, using 4 samples per iteration. In the second step, we run McMC with 4 chains, each sampling 500,000 samples. The total number of forward simulations was therefore 2 million, although this might be able to be decreased by employing variational inference methods for the second step or potentially by using more efficient Monte Carlo methods \cite{zunino2023hmclab}. For the two-step rj-McMC, 16 Markov chains with 3 million samples per chain were used for each period in the first-step, and 16 chains with 2 million samples per chain are used for each geographical location in the second-step, as reported in \citet{galetti2017transdimensional}; large numbers of samples were employed to ensure the convergence of the Markov chains due to the use of the trans-dimensional and hierarchical Bayesian approaches.

Note that in our current work we use a 2-step forward simulation method \cite{zhang20183}, which is based on similar approximations and assumptions to those made in the two-step inversion scheme: that the dispersion properties of surface waves at each location depend only on the velocity structure beneath that point. Ideally, we should use a fully 3D surface wave simulation method to account for any uncertainties and inaccuracy raised by this physical approximation of surface wave propagation, but this would increase the computational cost significantly. Future work should explore the extent to which this forward simulation method affects the inversion results.

In both the direct-3D inversion and standard two-step methods, our use of the same two-step forward simulation means that the same amount of computation is used to perform each simulation. Therefore, the main factor driving the difference in computational cost is the higher dimensionality of the direct-3D inversion compared to the two-step MH-McMC method, which makes the former extremely expensive to solve using conventional Monte Carlo sampling methods. We therefore used variational inference in our direct-3D inversion which reduces the computations significantly. Nevertheless, recent advances have introduced more efficient solutions for high-dimensional Bayesian inverse problems, such as neural network inversion \cite{earp2019probabilistic, zhang2021bayesian, hansen2022use, bloem2022introducing, scheiter2022upscaling, liu2024geostatistical, fone2024ambient, sun2024invertible}, gradient-based Monte Carlo sampling \cite{fichtner2019hamiltonian, gebraad2020bayesian, zhao2021gradient}, and variational inference as considered herein.

As stated in Section \ref{sec:inversion_setup}, main hyper-parameters (e.g., the prior bounds, likelihood function and the forward function) are set to be the same for all methods tested. Furthermore, the same dataset is used for all inversions, which should inject the same information, including lateral correlations, into the inversion results. This is the reason that the horizontal slices in Figure \ref{fig:uk_horizontal_mean_3methods} show consistent results in overview. Given these settings, a signifiant difference between the direct-3D and two-step methods is that the former is able to provide more accurate lateral correlation information in the inversion results, as presented in Figure \ref{fig:uk_vertical}. We note that significantly different vertical slices, but apparently consistent horizontal slices of the inversion results obtained using direct-3D rj-McMC and two-step rj-McMC methods were presented previously in \citet{zhang20183}. These differences are therefore due to the use of different inversion methods. In addition, Figure \ref{fig:uk_data_fit} shows that the direct-3D method provides more accurate inversion results with better data fit, compared to the two-step method. These results highlight the superiority of using a direct-3D method.

In addition, in Figures \ref{fig:uk_interpretation}b and \ref{fig:uk_interpretation}c we observe clear evidence of the Great Glen Fault, which is not revealed in the two-step rj-McMC results in Figure \ref{fig:uk_interpretation}d, nor in previous seismic imaging studies around the British Isles \cite{nicolson2014rayleigh, galetti2017transdimensional}. Since the data sets were identical, this can be attributed to the different inversion methods used. Previous studies imposed smoothing prior information into the inversion, either explicitly through regularisation terms using a linearised inversion \cite{nicolson2014rayleigh, bonadio2021optimal}, or implicitly via a trans-dimensional rj-McMC inversion approach using Voronoi cell parametrisation to reduce the dimensionality of the inverse problem \cite{galetti2017transdimensional}. These methods, while effective in some respects, tend to obscure finer structural details. This foregrounds why we use a regular gridded parametrisation (with a large number of cells, thus high dimensionality) for Bayesian 3D surface wave inversion: within a Bayesian framework we can remove any regularisation term safely without worrying about the problem of stability, such that in principle one find all possible model solutions that fit observed data to within estimated data uncertainties.

In this study, we tested the proposed direct-3D surface wave inversion method and constructed 3D seismic velocity models of the upper crust beneath Great Britain using Love wave group velocity dispersion data estimated from ambient noise cross-correlation. This is because Love wave group velocity maps are expected to be more representative of shallow sedimentary and superficial layers \cite{galetti2017transdimensional}. In addition, this removes the possibility that differences between the new results and those from \citet{galetti2017transdimensional} were due to the use of different data. Nevertheless, Love wave measurements are also known to be more challenging to measure than Rayleigh wave data because seismic data recorded on horizontal components contain stronger noise than data from the vertical component seismograms, and are more difficult to separate from overtones than Rayleigh wave data. Future work could focus on 3D joint inversion using different types of geophysical data to build more accurate subsurface models beneath Great Britain \cite{arrowsmith2005seismic, nicolson2014rayleigh, galetti2017transdimensional, montiel2025first}.

\section{Conclusion}
We perform Bayesian surface wave inversion across Great Britain and present the first high resolution seismic velocity models of the upper crust using Love wave dispersion data estimated from ambient noise cross-correlation. We introduce a new, variational, direct-3D inversion method, and compare it to two-step MH-McMC and rj-McMC methods. The results from the direct-3D method exhibit desirable lateral correlations which are absent in the results produced by the two-step methods. In addition, when comparing observed data with synthetic data simulated from the direct-3D and two-step MH-MCMC inversion results, the former is proven to be more accurate. These models provide significant insights into the crustal structure of the region, and accurately reveal several well-known geological features, such as the Great Glen Fault. Based on our findings, we recommend that the seismology community should prioritise fully 3D inversion methods for regional surface wave inversion.

\section{ACKNOWLEDGMENTS}
	We thank the Edinburgh Imaging Project sponsors (BP and TotalEnergies) for supporting this research. For the purpose of open access, we have applied a Creative Commons Attribution (CC BY) licence to any Author Accepted Manuscript version arising from this submission.

\bibliographystyle{plainnat}  
\bibliography{reference}

\appendix
\section{More Inversion Results}
\label{ap:inversion_results}

\begin{figure}
	\centering\includegraphics[width=\textwidth]{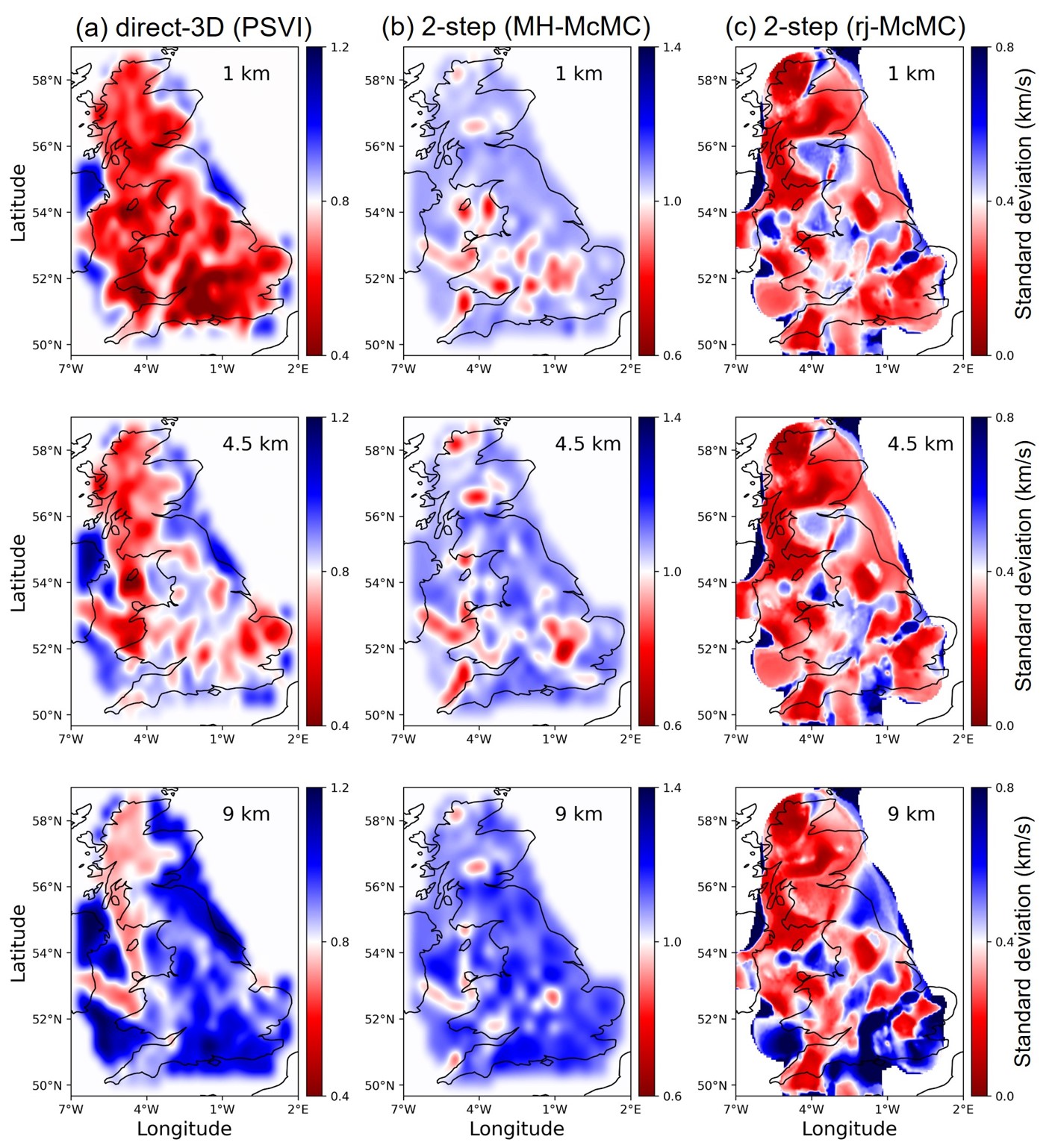}
	\caption{Posterior standard deviation maps as in Figure \ref{fig:uk_horizontal_std_3methods_samescale} in the main text, but using different color scales to highlight different uncertainty structures from different methods. Key as in Figure \ref{fig:uk_horizontal_std_3methods_samescale}.}
	\label{fig:uk_horizontal_std_3methods}
\end{figure}

\begin{figure}
	\centering\includegraphics[width=\textwidth]{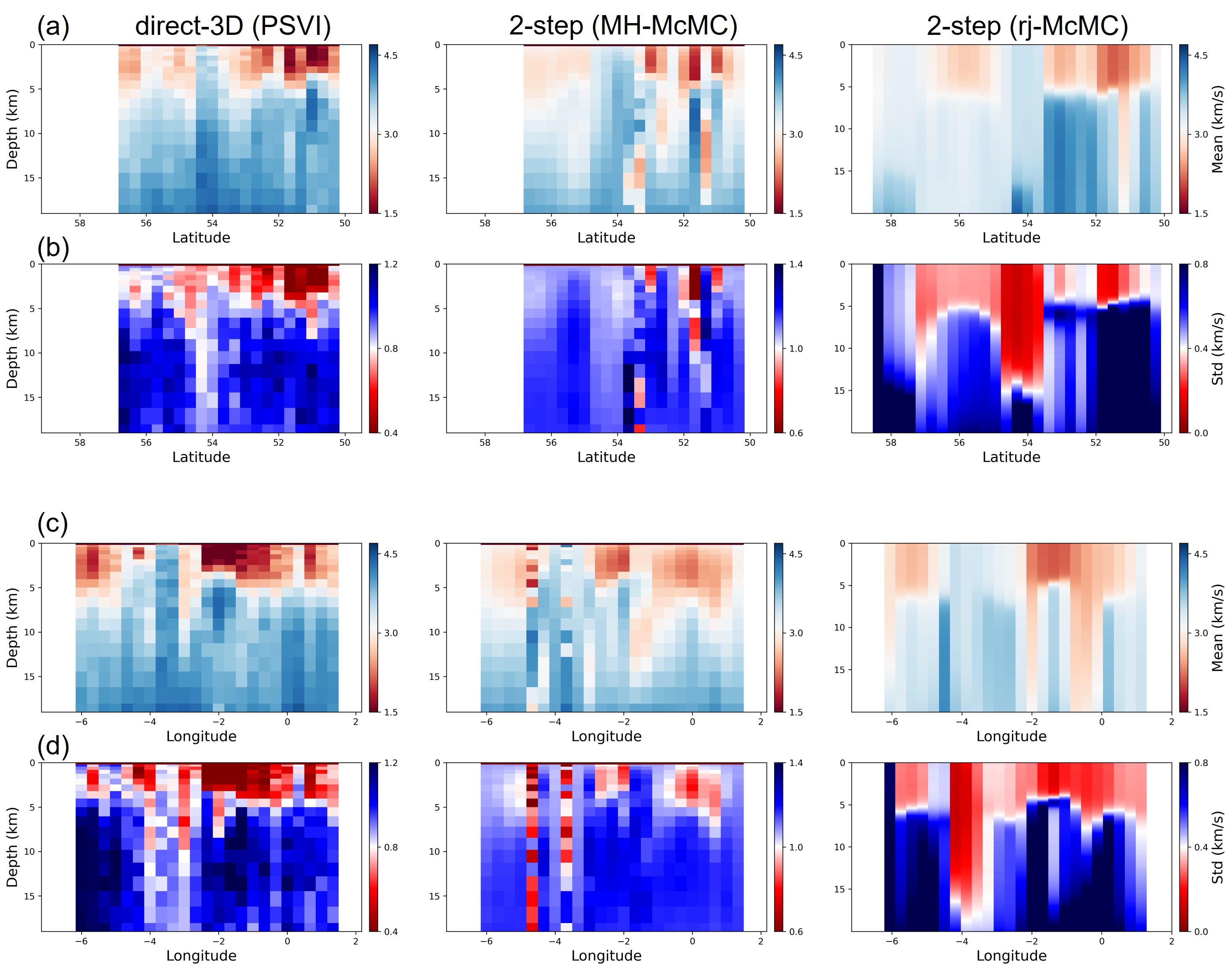}
	\caption{Vertical slices of the inversion results as in Figure \ref{fig:uk_vertical} in the main text, but the standard deviation maps are displayed with different color scales for different methods. Key as in Figure \ref{fig:uk_vertical}.}
	\label{fig:uk_vertical_diffscale}
\end{figure}

In this Appendix, we provide more inversion results. Figures \ref{fig:uk_horizontal_std_3methods} and \ref{fig:uk_vertical_diffscale} show the inversion results in which different color scales are used for the standard deviation maps from different methods, to better illustrate detailed uncertainty structures of the three sets of inversion results.

\begin{figure}
	\centering\includegraphics[width=\textwidth]{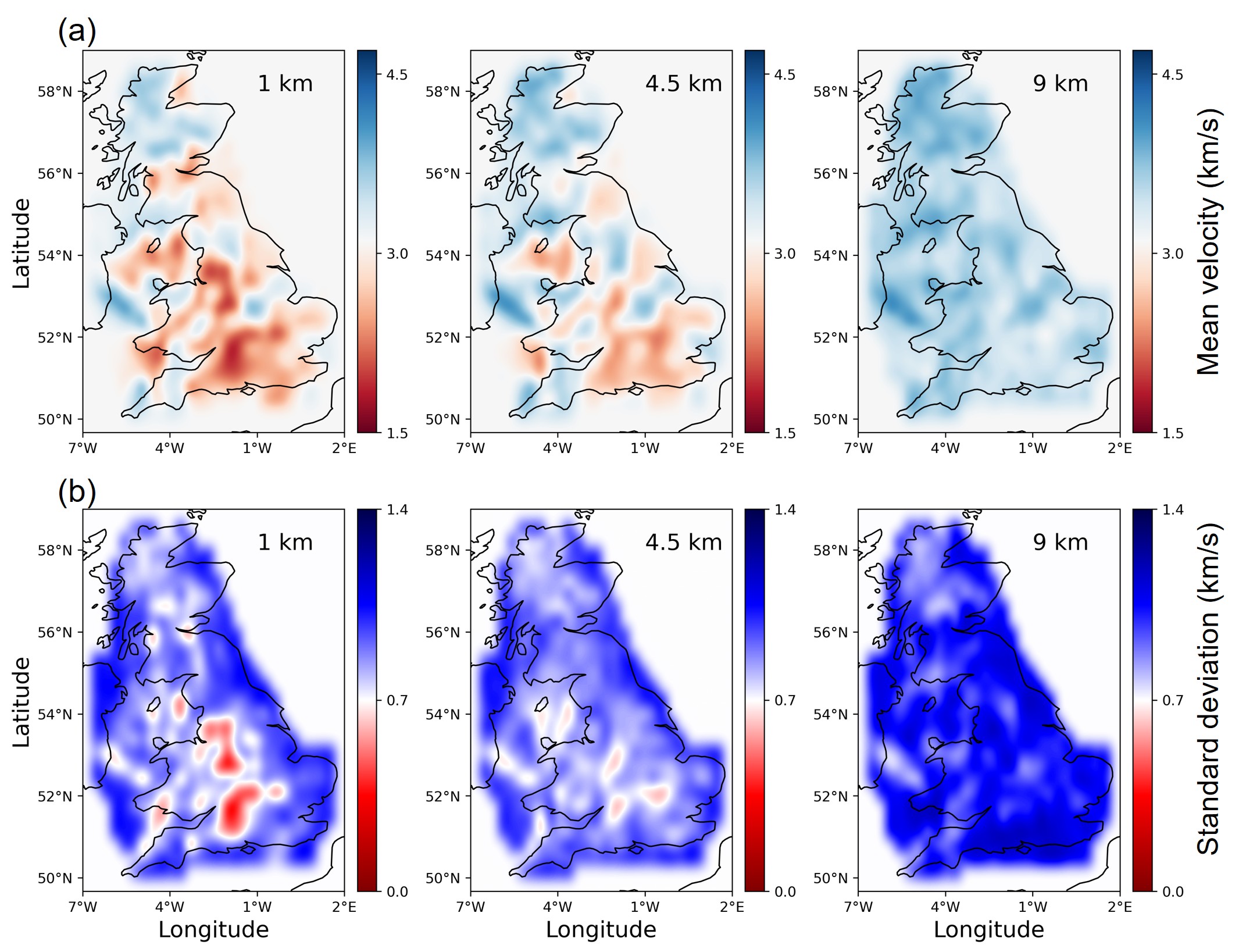}
	\caption{Three horizontal slices of the inversion results obtained using the two-step PSVI method at depths of 1 km, 4.5 km and 9 km, in which both two separate inversions are performed using PSVI, for comparison with the two-step MH-McMC results presented in Figures \ref{fig:uk_horizontal_mean_3methods} and \ref{fig:uk_horizontal_std_3methods_samescale} in the main text. (a) Posterior mean and (b) standard deviation maps.}
	\label{fig:uk_horizontal_psvi_twostep}
\end{figure}

\begin{figure}
	\centering\includegraphics[width=\textwidth]{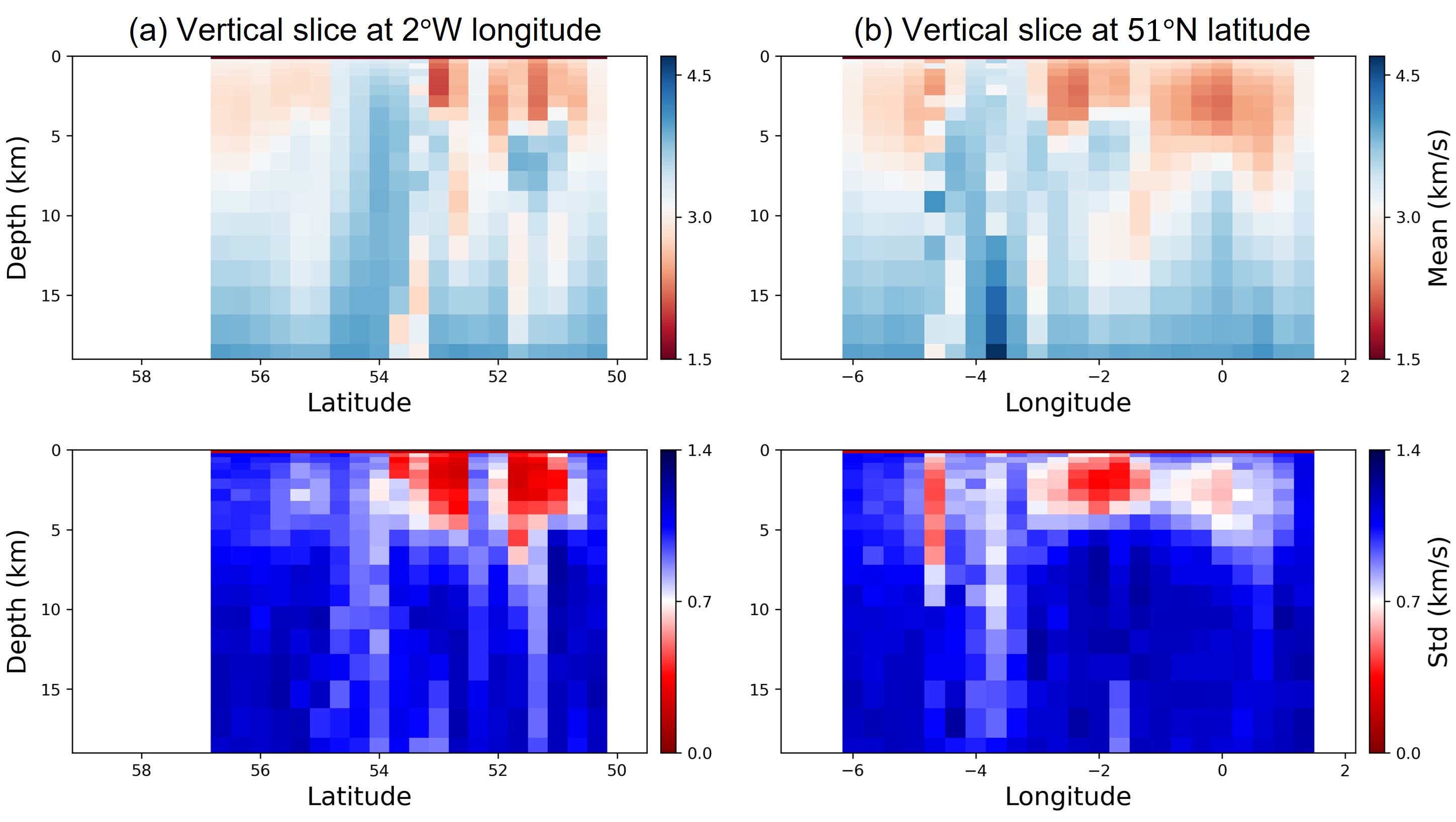}
	\caption{Two vertical slices of the inversion results obtained using the two-step PSVI method at (a) 2$^\circ$W longitude and (b) 51$^\circ$N latitude. Key as in Figure \ref{fig:uk_vertical} in the main text.}
	\label{fig:uk_vertical_psvi_twostep}
\end{figure}

Figures \ref{fig:uk_horizontal_psvi_twostep} and \ref{fig:uk_vertical_psvi_twostep} show the inversion results obtained using a two-step method with PSVI applied in both steps. The results are highly consistent with those from the two-step MH-McMC method displayed in Figures \ref{fig:uk_horizontal_mean_3methods}, \ref{fig:uk_horizontal_std_3methods_samescale} and \ref{fig:uk_vertical} in the main text, and thus prove that the main differences between the direct-3D and two-step MH-McMC inversion results are caused by splitting the inversion into two separate steps, rather than using different probabilistic algorithms such as MH-McMC versus PSVI in the second step (depth inversion).

\begin{figure}
	\centering\includegraphics[width=\textwidth]{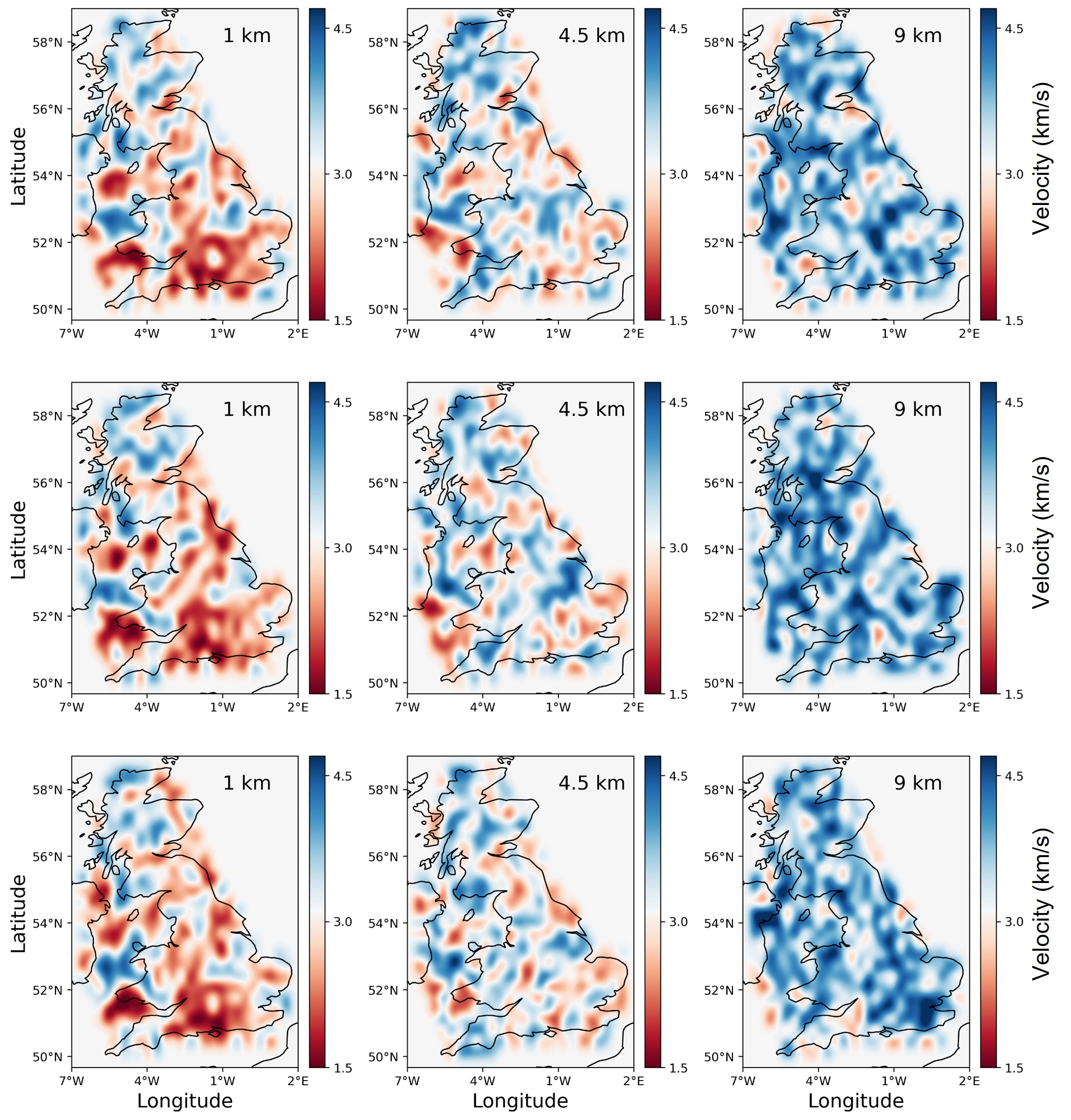}
	\caption{Horizontal slices of three posterior samples that have the highest likelihood values (thus the lowest data misfit values), obtained from the direct-3D inversion at depths of 1 km, 4.5 km and 9 km. Each row shows one posterior sample.}
	\label{fig:uk_horizontal_3d_bestfit_samples}
\end{figure}

\begin{figure}
	\centering\includegraphics[width=\textwidth]{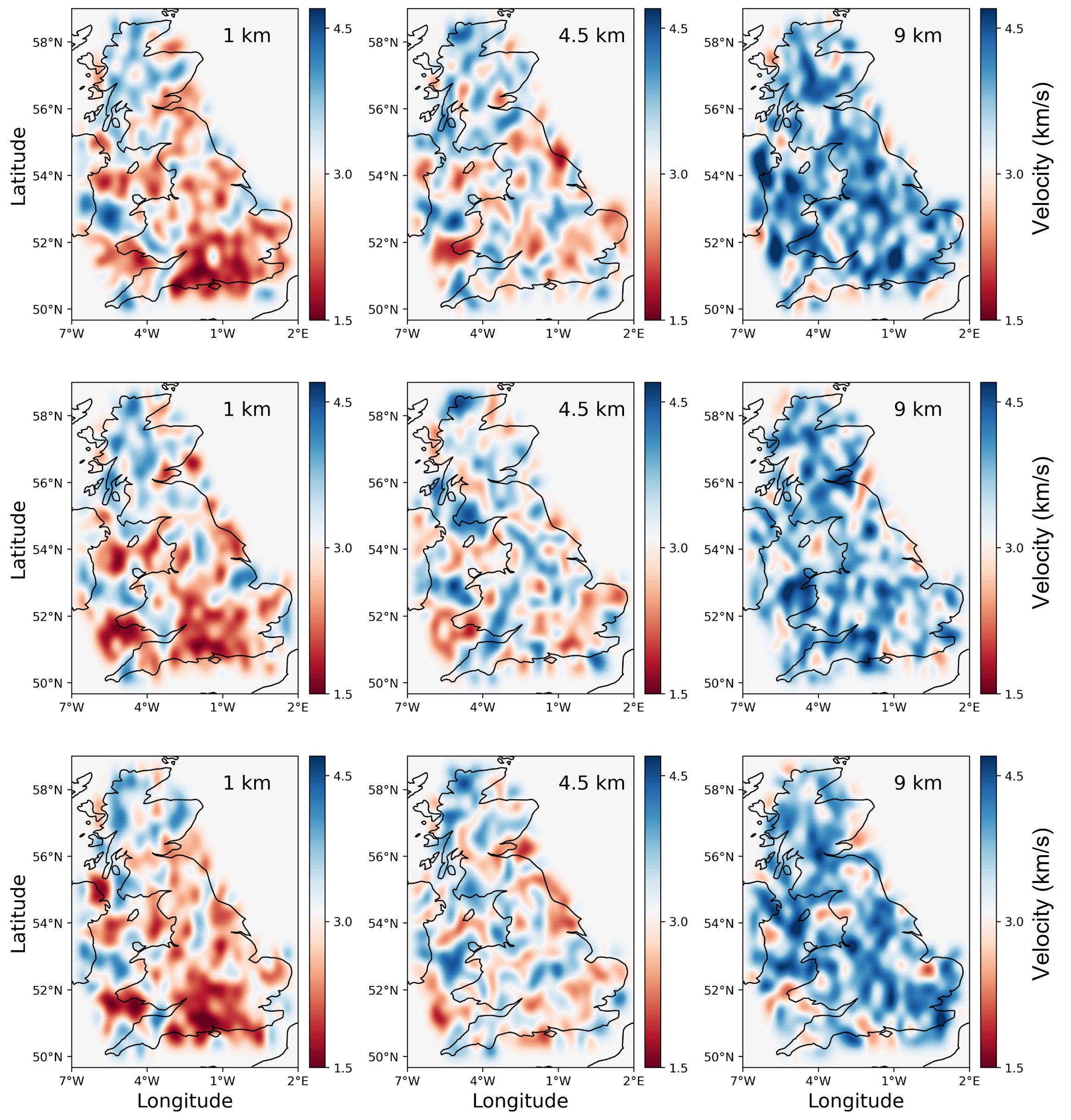}
	\caption{Horizontal slices of three posterior samples that have the lowest likelihood values (thus the highest data misfit values), obtained from the direct-3D inversion at depths of 1 km, 4.5 km and 9 km. Each row shows one posterior sample.}
	\label{fig:uk_horizontal_3d_worstfit_samples}
\end{figure}

To present the nonuniqueness property of the inversion results obtained from the direct-3D inversion, in Figures \ref{fig:uk_horizontal_3d_bestfit_samples} and \ref{fig:uk_horizontal_3d_worstfit_samples} we show two sets of posterior samples. Figure \ref{fig:uk_horizontal_3d_bestfit_samples} isolates models with the highest likelihood values (thus the lowest data misfit values), and Figure \ref{fig:uk_horizontal_3d_worstfit_samples} shows models with the lowest likelihood values. Although these two sets of samples look different to some extent, they all present similar features to the three posterior mean models displayed in Figure \ref{fig:uk_horizontal_mean_3methods} in the main text, since all posterior samples are expected to fit observed data to within data uncertainties.

%
%

\begin{figure}
	\centering\includegraphics[width=\textwidth]{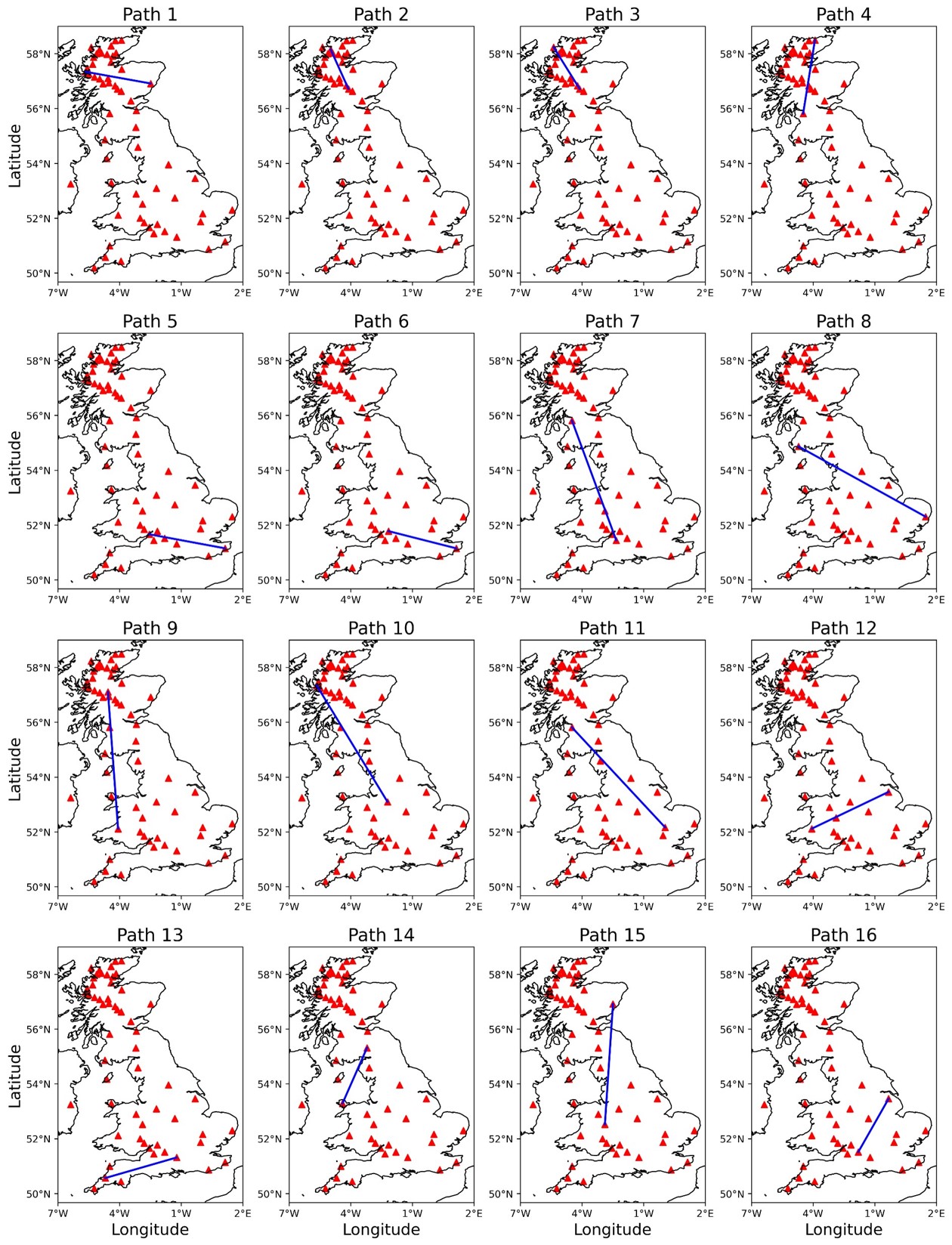}
	\caption{16 inter-receiver ray paths used to compare data fit in Figures \ref{fig:uk_16paths_data_3d} and \ref{fig:uk_16paths_data_1d}.}
	\label{fig:uk_16paths}
\end{figure}

\begin{figure}
	\centering\includegraphics[width=\textwidth]{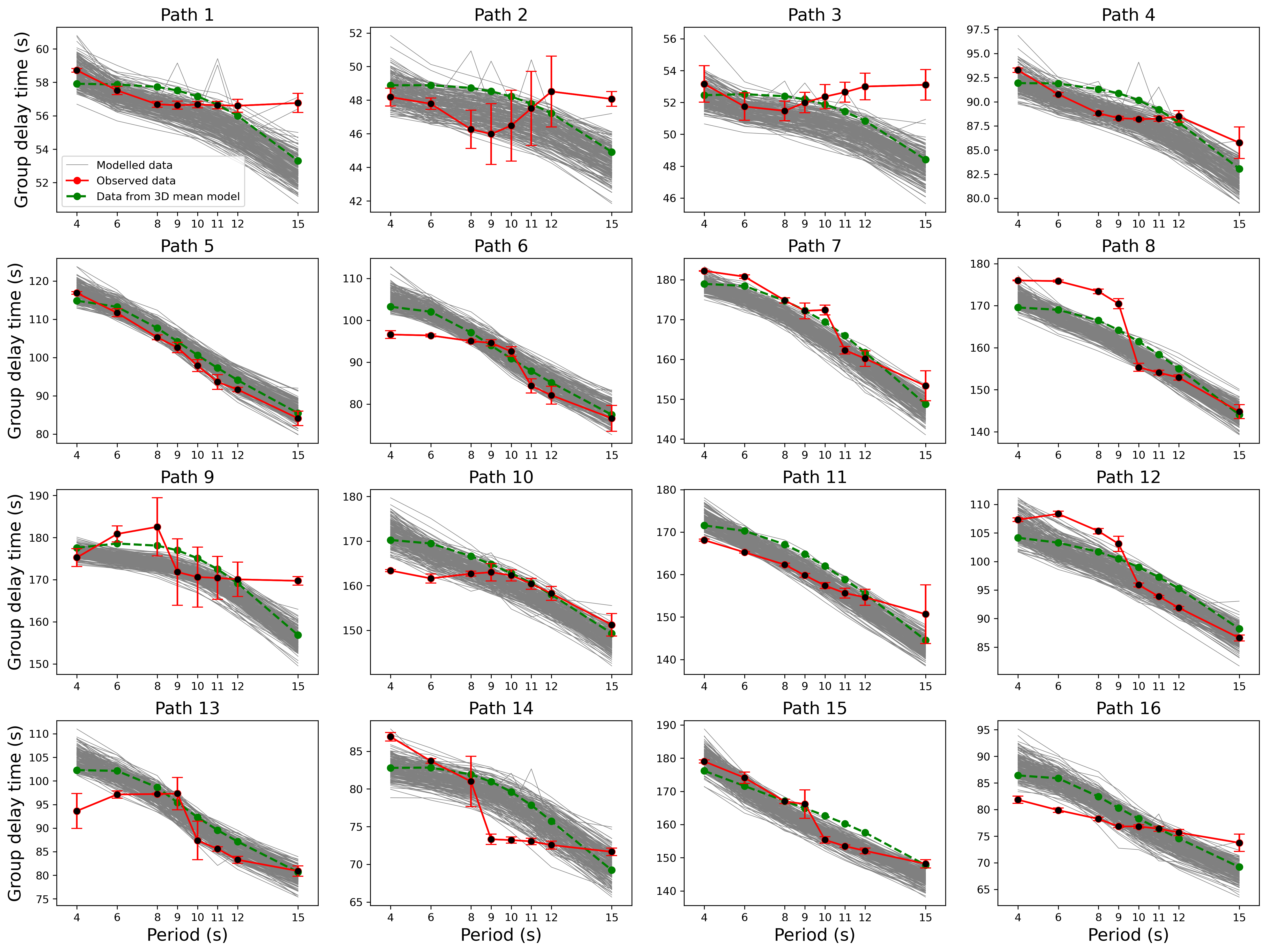}
	\caption{Synthetic Love wave group delay times (grey lines) across 16 inter-receiver ray paths (Figure \ref{fig:uk_16paths}) at 8 periods, obtained using posterior samples of the direct-3D inversion results. Red lines stand for observed data, and dashed green lines stand for data simulated from the inverted mean velocity model.}
	\label{fig:uk_16paths_data_3d}
\end{figure}

\begin{figure}
	\centering\includegraphics[width=\textwidth]{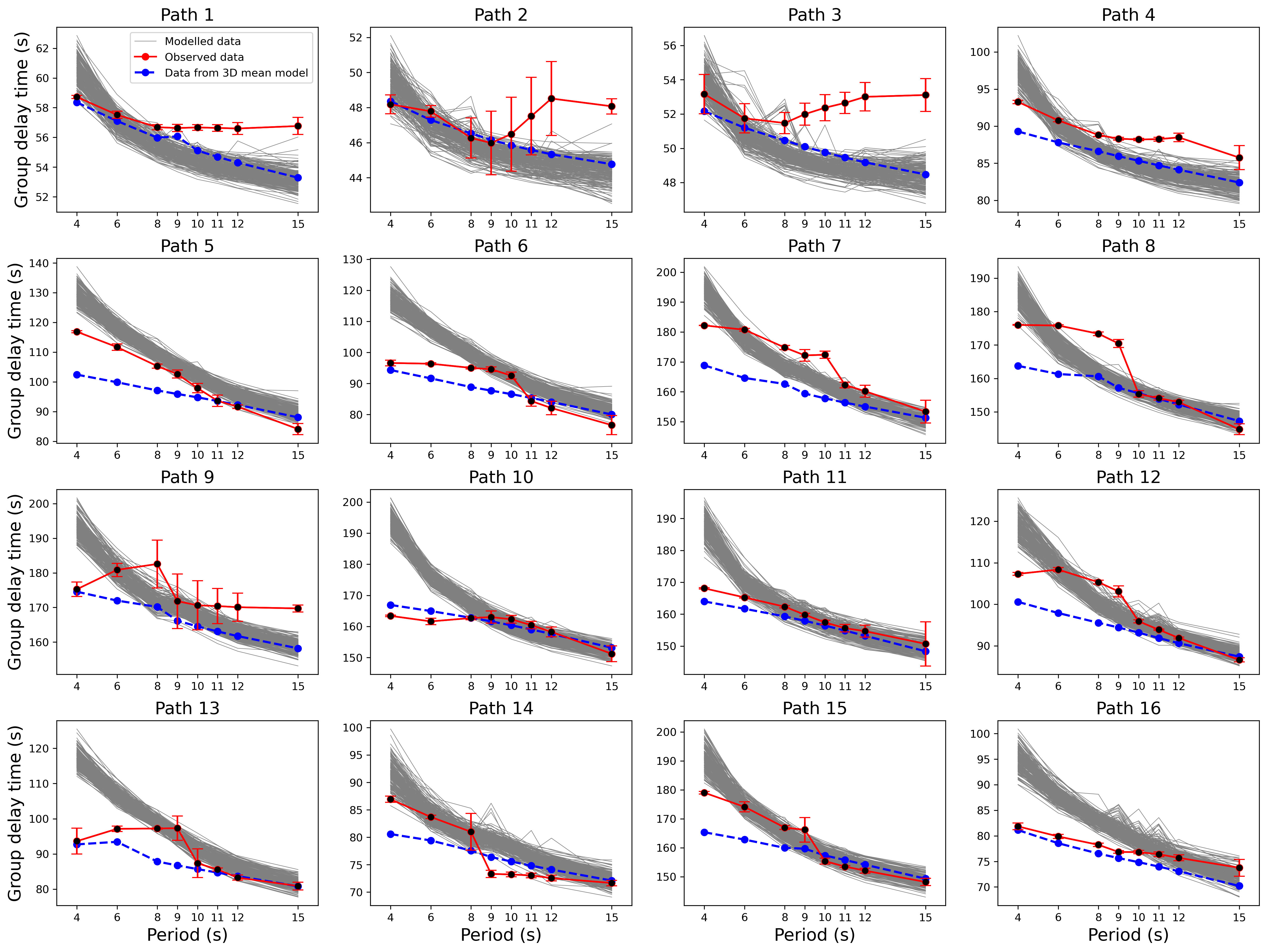}
	\caption{Synthetic Love wave group delay times simulated using posterior samples from the two-step MH-McMC inversion results. Key as in Figure \ref{fig:uk_16paths_data_3d}.}
	\label{fig:uk_16paths_data_1d}
\end{figure}

To further support the statement that the direct-3D method yields more accurate inversion results and superior data fitting compared to the two-step MH-McMC method, we analysed an additional 16 inter-receiver ray paths, as displayed in Figure \ref{fig:uk_16paths}. Figures \ref{fig:uk_16paths_data_3d} and \ref{fig:uk_16paths_data_1d} show synthetic Love wave group delay time data between these 16 ray paths, simulated using posterior samples from the direct-3D and two-step MH-McMC inversion results, respectively, similarly to those displayed in Figures \ref{fig:uk_data_fit}a and \ref{fig:uk_data_fit}b in the main text. Once again, the synthetic data produced by the direct-3D inversion method demonstrate better fit to the observed data compared to those generated by the two-step MH-McMC method, thus reinforcing the conclusions drawn in the main text.

\section{An Additional Synthetic Example}
\label{ap:synthetic}

In this section, we compare the direct-3D and two-step MH-McMC inversion methods using a synthetic checkerboard example that mirrors the setup of our real data example. This includes employing the same station locations, defining the same inversion region (bounded by longitude 7$^\circ$W-2$^\circ$E and latitude 49.6$^\circ$N-59$^\circ$N), and using the same regular-gridded parametrisation. The only distinction is in the \textit{true} velocity model used. 

\begin{figure}
	\centering\includegraphics[width=\textwidth]{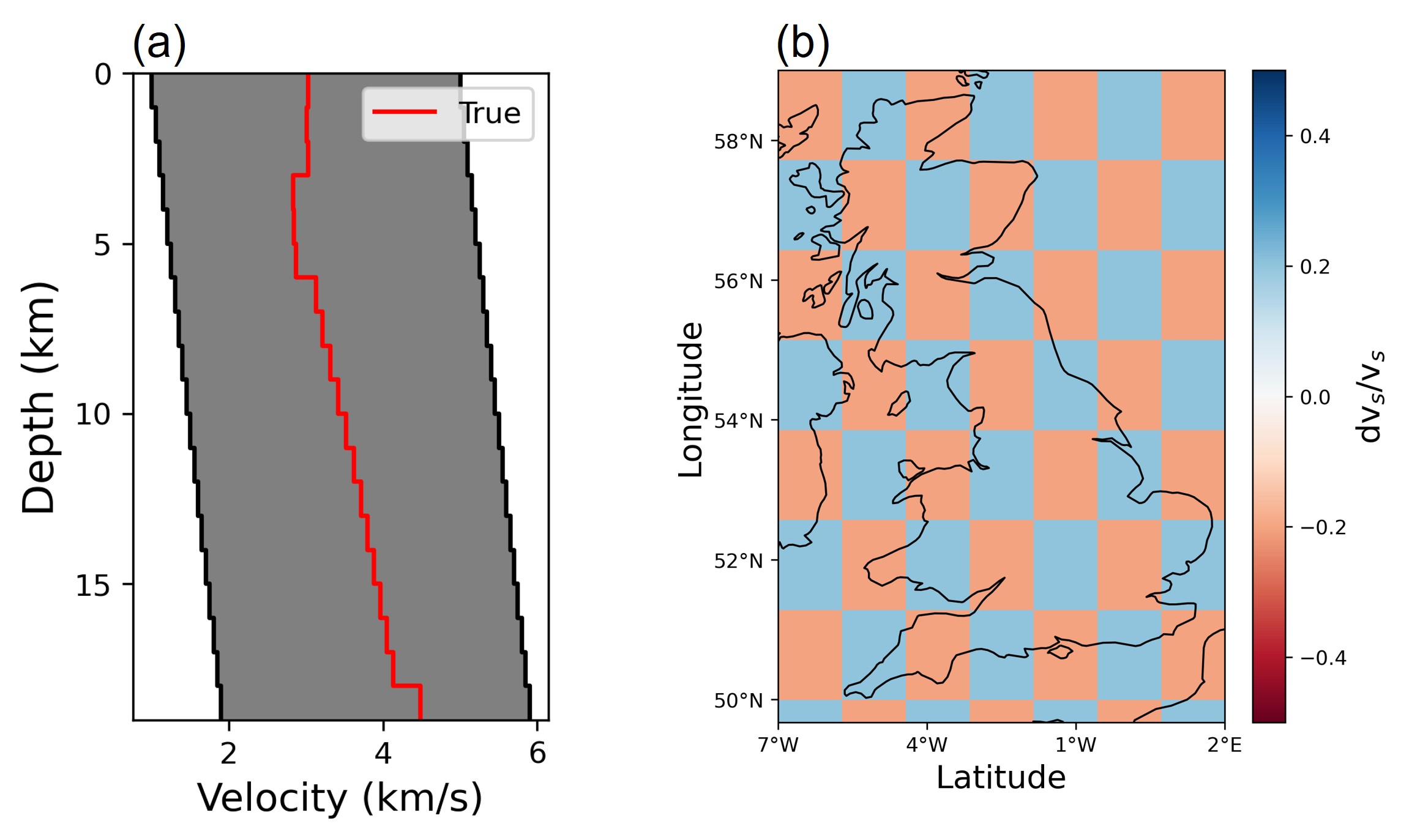}
	\caption{(a) Background true shear wave velocity profile (red line) at different depths. Grey area defines the range of a uniform prior pdf used in this example. (b) A checkerboard velocity pattern used to define a 3D velocity model.}
	\label{fig:synthetic_prior}
\end{figure}

Initially, we define a laterally homogeneous, 20-layered 3D shear wave velocity model. The velocities for different layers are depicted by a red line in Figure \ref{fig:synthetic_prior}a. Within each layer, we perturb the true velocity by $\pm$20\% to generate a checkerboard pattern, illustrated in Figure \ref{fig:synthetic_prior}b. This 3D velocity model is then used to compute Love wave measurements across 8 periods identical to those used in the real data example, serving as the observed dataset in this synthetic scenario.

\begin{figure}
	\centering\includegraphics[width=\textwidth]{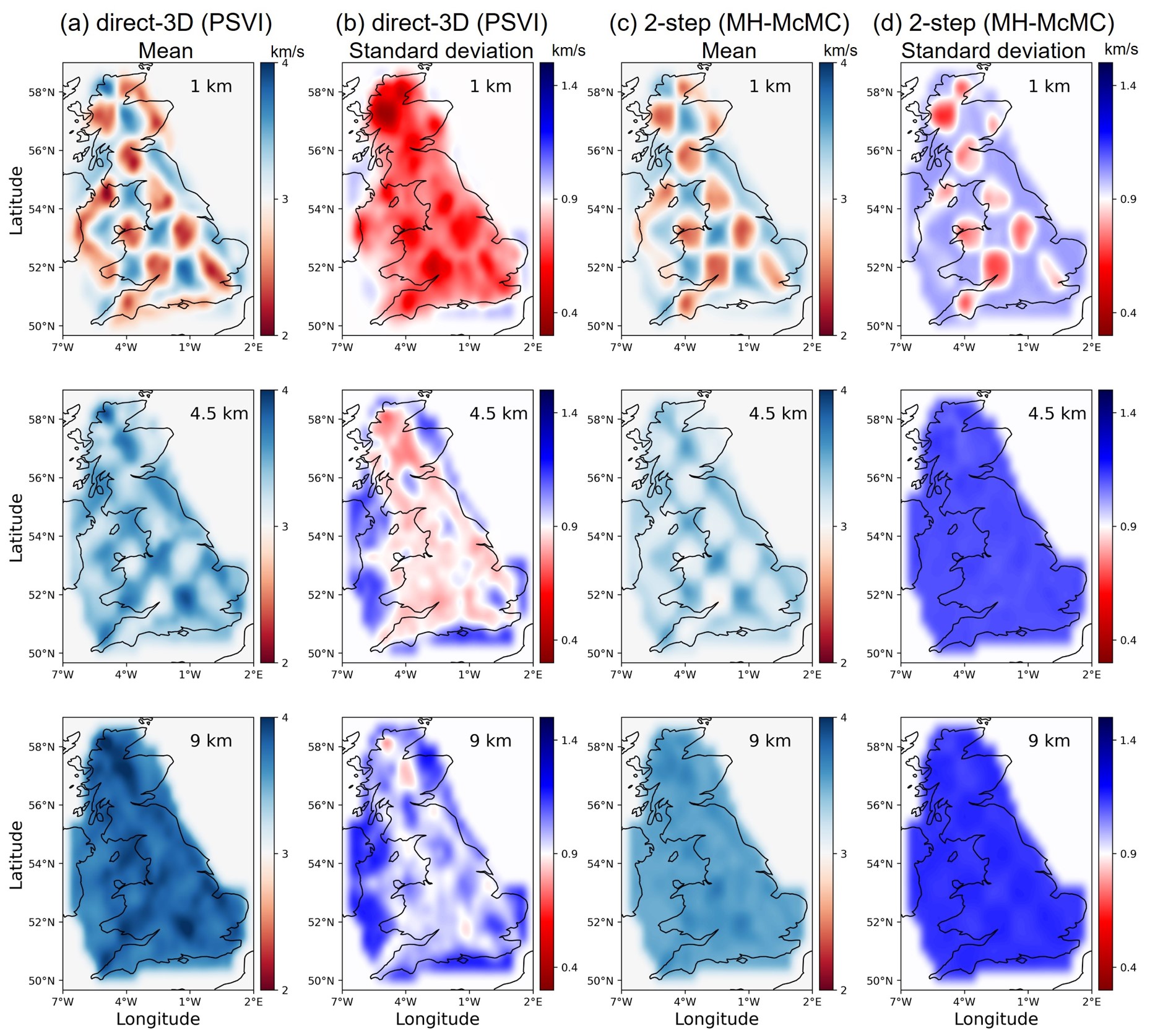}
	\caption{Horizontal slices of the inversion results (shear wave velocity structures) at depths of 1 km, 4.5 km, and 9 km, respectively. (a) Mean and (b) standard deviation maps from the direct-3D inversion; (c) and (d) are those from the two-step MH-McMC inversion.}
	\label{fig:synthetic_horizontal}
\end{figure}

We employ both the direct-3D and two-step MH-McMC inversion methods and use the same hyper-parameters as those detailed in the main text. Figure \ref{fig:synthetic_horizontal} displays three horizontal slices of the inversion results. Similarly to our previous findings, the two sets of mean velocity maps are consistent, yet the posterior standard deviation values from the two-step MH-McMC inversion are larger than those from the direct-3D inversion.

\begin{figure}
	\centering\includegraphics[width=\textwidth]{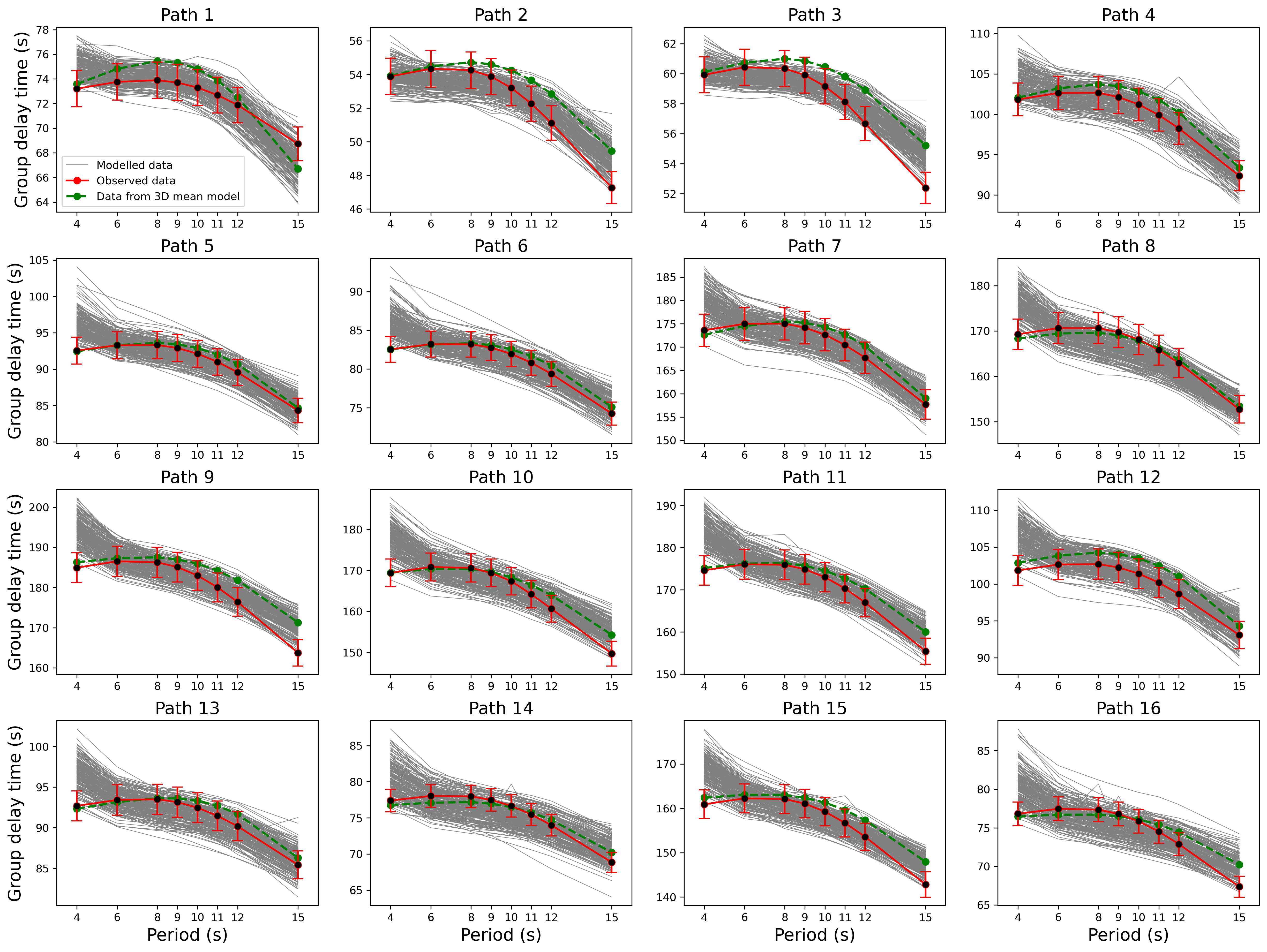}
	\caption{Synthetic Love wave group delay times simulated using posterior samples from the direct-3D inversion results. Key as in Figure \ref{fig:uk_16paths_data_3d}.}
	\label{fig:synthetic_16paths_data_3d}
\end{figure}

\begin{figure}
	\centering\includegraphics[width=\textwidth]{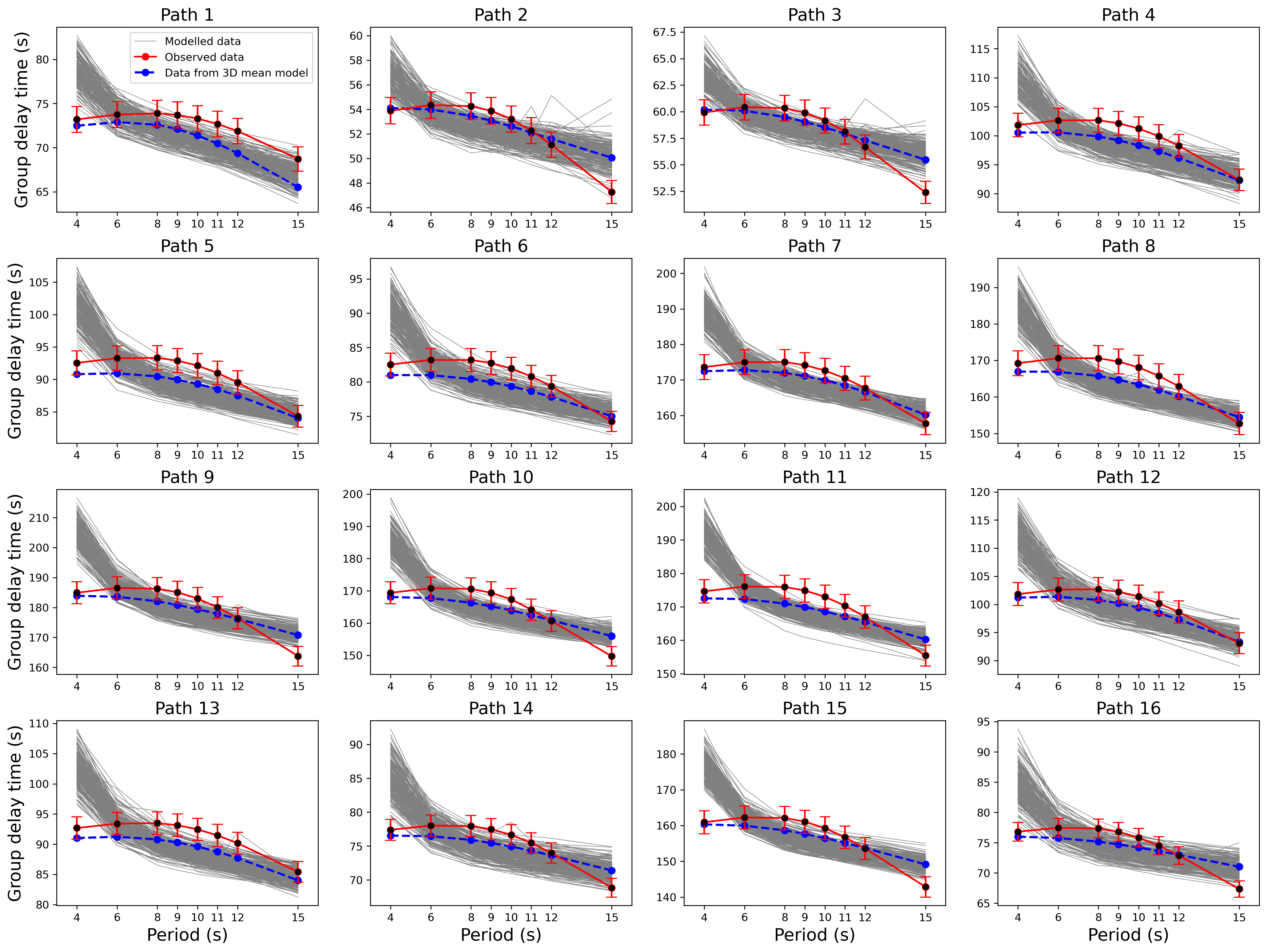}
	\caption{Synthetic Love wave group delay times simulated using posterior samples from the two-step MH-McMC inversion results. Key as in Figure \ref{fig:uk_16paths_data_1d}.}
	\label{fig:synthetic_16paths_data_1d}
\end{figure}

We then conduct an identical test to compare observed data, and data simulated from the two sets of inversion results. Figures \ref{fig:synthetic_16paths_data_3d} and \ref{fig:synthetic_16paths_data_1d} illustrate synthetic Love wave group delay times calculated using posterior samples from the direct-3D and two-step MH-McMC inversion methods, respectively, across the same 16 inter-receiver ray paths shown in Figure \ref{fig:uk_16paths}. Furthermore, Figure \ref{fig:synthetic_data_fit} in the main text compares normalized data misfit values and logarithmic likelihood values for the 8 periods, similarly to Figure \ref{fig:uk_data_fit} in the main text.

Again, this synthetic example demonstrates that the inversion results from the direct-3D method are more accurate than those from the two-step MH-McMC method, as indicated by the closer fit of the synthetic data to the observed data in the former. Consequently, this reaffirms that the conclusions drawn in the main text are broadly applicable to seismic surface wave inversion problems.

\label{lastpage}
\end{document}